\documentclass[useAMS,usenatbib]{mn2e}
\def\spose#1{\hbox to 0pt{#1\hss}}
\def\approxlt{\mathrel{\spose{\lower 3pt\hbox{$\sim$}}
	\raise 2.0pt\hbox{$<$}}}
\def\approxgt{\mathrel{\spose{\lower 3pt\hbox{$\sim$}}
	\raise 2.0pt\hbox{$>$}}}
\def\approxpropto{\mathrel{\spose{\lower 3pt\hbox{$\sim$}}
	\raise 2.0pt\hbox{$\propto$}}}
\mathchardef\twiddle="2218

\def\multleft#1{\hbox to size{\vbox {\halign {\lft{##}\cr #1}}\hfill}\par}
\def\multright#1{\hbox to size{\vbox {\halign {\rt{##}\cr #1}}\hfill}\par}

\def\today{\ifcase\month\or January\or February\or March\or April\or May\or
      June\or July\or August\or September\or October\or November\or December\fi
      \space\number\day, \number\year}
\def\<{\thinspace}

\def\arcsec{{\rm\thinspace arcsec}}
\def\cm{{\rm\thinspace cm}}
\def\erg{{\rm\thinspace erg}}

\def\keV{{\rm\thinspace keV}}

\def\km{{\rm\thinspace km}}
\def\kpc{{\rm\thinspace kpc}}

\def\Mpc{{\rm\thinspace Mpc}}

\def\s{{\rm\thinspace s}}
\def\yr{{\rm\thinspace yr}}

\usepackage{graphicx,graphics}

\usepackage{txfonts}
\usepackage{psfig}
\usepackage{epsfig}
\usepackage{natbib}
\usepackage{wrapfig}
\usepackage{amssymb}
\usepackage{url}
\usepackage{longtable}
\usepackage{times}
\usepackage{float}
\usepackage[usenames]{color}
\usepackage{soul,color}
\usepackage{subfigure}
\newcommand*{\mysub}[2]{\ensuremath{#1_{\mathrm{#2}}}}
\newcommand*{\Omegam}{\mysub{\Omega}{m}}

\newcommand*{\Omegal}{\ensuremath{\Omega_{\Lambda}}}

\newcommand*{\LCDM}{\ensuremath{\Lambda}CDM}

\newcommand*{\msolar}{\mysub{M}{\odot}}
\newcommand*{\zsolar}{\mysub{Z}{\odot}}

\title[ Extreme Feedback in \MACS]
  {  Extreme AGN Feedback and Cool Core Destruction in the X-ray Luminous Galaxy Cluster MACS J1931.8-2634}
\author[S. Ehlert et al.]
  {S.~Ehlert,$^1$\thanks{Email:sehlert@stanford.edu}
  S.W.~Allen,$^1$ 
  A.~von der Linden,$^1$ 
  A.~Simionescu,$^1\thanks{Einstein Fellow}$ 
  N.~Werner,$^1\thanks{Chandra/Einstein Fellow}$
  G.B.~Taylor,$^2$
  \newauthor 
  G.~Gentile,$^3$
  H.~Ebeling,$^4$
  M.T.~Allen,$^5$
  D.~Applegate,$^1$
  R.J.H.~Dunn,$^6$
  A.C.~Fabian,$^7$
  \newauthor
  P.~Kelly,$^1$
  E.T.~Million,$^1$
  R.G.~Morris,$^1$
  J.S.~Sanders,$^7$ 
  and R.W.~Schmidt$^8$ \\ 
  $^1$Kavli Institute for Particle Astrophysics and Cosmology at Stanford University, 452 Lomita Mall, Stanford, CA 94305-4085, USA\\
  and SLAC National Accelerator Laboratory, 2575 Sand Hill Road, Menlo Park, CA 94025, USA\\
  $^2$University of New Mexico, Department of Physics and Astronomy, Alberquerque, NM 87131, USA\\
  Greg Taylor is also an Adjunct Astronomer at the National Radio Astronomy Observatory\\
  $^3$Sterrenkundig Observatorium, Universiteit Gent, Krijgslaan 281, B-9000 Gent, Belgium\\
  $^4$Institute for Astronomy, University of Hawaii, 2680 Woodlawn Drive, Honolulu, HI 96822, USA\\
  $^5$Stanford University, Stanford, California 94305-4060, USA\\
  $^6$Excellence Cluster `Universe', Technische Universit\"at M\"unchen, Boltzmannstrasse 2, D-85748, Garching, Germany\\
  $^7$Intitute of Astronomy, Madingley Road, Cambridge, CB3 0HA\\
  $^8$Astronomisches Rechen-Institut, Zentrum f\"ur Astronomie der Universit\"at Heidelberg, M\"onchhofstrasse 12-14, 69120 Heidelberg, Germany}

\def\cha{{\it Chandra}}
\def\MACS{{MACS J1931.8-2634}}
\def\sub{{\it Subaru}}
\def\vla{{\it VLA}}
\def\mach{{\mathcal{M} }}
\def\halpha{H$\alpha$ }
\def\msolaryr{ $\msolar \yr^{-1}$}
\def\ergs{$\erg \s^{-1}$}
\def\h70{\mysub{h}{70}}

\def\arcdeg {\hbox{$^{\circ}$}}
\def\arcsecf {\hbox{$.\!\!^{\prime\prime}$}} 
\def\arcminf {\hbox{$.\!\!^{\prime}$}}

\def\arcsecf {\hbox{$.\!\!^{\prime\prime}$}} 
\def\arcminf {\hbox{$.\!\!^{\prime}$}}
\def\degree {\hbox{$^{\circ}$}}

\begin{document}

\date{Accepted 2010 September 29. Received 2010 August 24; in original form 2010 June 21}

\maketitle

\begin{abstract}
We report on a deep, multiwavelength study of the galaxy cluster \MACS \ using \cha \ X-ray, \sub \ optical, and \vla \ 1.4 GHz radio data.
This cluster ($z=0.352$) harbors one of the most X-ray luminous cool cores yet discovered, 
with an equivalent mass cooling rate within the central $50\h70^{-1} \kpc$ \ is $\sim$700 \msolaryr. Unique
features observed in the central core of \MACS \  hint to a wealth of past activity that has greatly disrupted the original cool core.
The X-ray and optical data suggest oscillatory motion of the cool core along a roughly north-south direction. We 
also observe a spiral of relatively cool, dense, X-ray emitting gas connected to the cool core, as well as highly 
elongated intracluster light (ICL) surrounding the cD galaxy. For a cluster with such a high nominal cooling rate, this cluster is missing the central metallicity peak
almost always seen in cool core clusters, which suggests bulk transport of cool gas out to large distances from the center. 
Extended radio emission is observed surrounding the 
central AGN, elongated in the east-west direction, spatially coincident with X-ray cavities. The
power input required to inflate these `bubbles' is estimated from both the X-ray 
and radio emission to reside between $\mysub{P}{jet} \sim$4 -- 14 $\times 10^{45}$ \ergs, 
putting it among the most powerful jets ever observed.    
This combination of a  powerful AGN outburst and  bulk motion of the cool core have resulted in two X-ray bright ridges 
to form to the north and south of the central AGN at a distance of approximately 
25 \kpc. The northern ridge has spectral characteristics typical of cool cores: it contains low temperature, high density, metal rich gas
and is consistent with being a remnant of the cool core after it was disrupted by the AGN and bulk motions. It is also the site of \halpha filaments and young stars.
The X-ray spectroscopic cooling rate associated with this
ridge is $\sim$165 \msolaryr, which agrees with the estimate of the star formation rate from broad-band optical imaging ($\sim$170 \msolaryr). 
\MACS \ appears to harbor one of 
most profoundly disrupted low entropy cores observed in a cluster, and offers new insights into the 
survivability of cool cores in the context of hierarchical structure formation. 

 \end{abstract}


\section{Introduction}\label{intro}
The relatively cool and dense gas at the centers of many galaxy clusters emits copious X-ray radiation by thermal 
bremsstrahlung and line emission. In the absence of
external sources of heating, this high emission should lead to very rapid cooling ($\tau_{cool} < 1 \ \rm{Gyr}$) 
and very high rates of mass deposition onto the central
cluster galaxy (up to $\sim$1000 \msolaryr), in turn causing very high star formation rates and strong line 
emission around 0.5-1.0 \keV \ \citep[e.g.][]{Fabian1977,Cowie1977,Peterson2006,
McNamara2007}. The lack of such obvious observational signatures \citep[e.g.][]{Peterson2001,Peterson2003} provides compelling evidence 
that some central source of heating must be present. The most plausible source of heating is
feedback from the central Active Galactic Nucleus (AGN). 
Large X-ray cavities filled with radio plasma are clearly seen in many systems, 
which provide an expected source of 
heating and turbulence \citep[e.g.][]{Brueggen2002}. In systems such as the Perseus, Virgo, 
Centaurus, and Hydra A Clusters \citep[e.g.][Million et al. submitted]{Fabian2003,Forman2005,Nulsen2005Hydra,
Fabian2006,Forman2007,Sanders2007,Sanders2008,Simionescu2009}, heating of the ICM has been argued to involve sound waves 
and weak shocks. Many questions remain as to the precise processes by which central AGN
activity suppresses cooling flows, and the 
extent to which these feedback processes require finely tuned parameters. More ambiguous still are the nature
and energetics of feedback mechanisms at higher redshifts, as only a few systems with such high classical cooling rates
have been studied in detail \citep{Schindler1997,Allen2000,Kitayama2004,Bohringer2005,Ogrean2010}.

The galaxy cluster \MACS \ is an extreme example of a cluster with a rapidly cooling core, making it an ideal 
system to test the limits of feedback mechanisms in 
galaxy clusters. In a short 12 ks \cha \ observation of \MACS \ taken in October of 2002 \citep{Allen2004,Allen2008}, X-ray cavities
 were detected surrounding a bright central 
AGN. The physical size of these cavities ($\sim$25 $\kpc$) is similar to those observed in the nearby 
Perseus Cluster \citep{Fabian2003,Fabian2006}. Indeed,
with its luminous cool core, visible central point source, and very large apparent cooling rate, \MACS \ 
is in many ways a higher redshift 
analog to the Perseus Cluster. Deeper observations of \MACS \ were
taken with 
\cha \ in August of 2008, increasing the total clean exposure to $\sim$100  $\rm{ks}$. 
The combined X-ray data are presented for the first time here, and are
complimented with optical (\sub) and radio (\vla) observations. Our goals are to acquire a better understanding of the thermodynamic structure around the
central AGN and the extent to which feedback from the central AGN counteracts this extreme cooling flow.

The structure of this paper is as follows: Section \ref{datared} discusses the reduction of the \cha \ data. 
Section \ref{imaging} discusses the imaging analysis of the X-ray data, while Section \ref{deprojanalysis} presents results
 on the mass profile and cooling flow of \MACS. Section \ref{spectra} discusses the spectral analysis.
Sections \ref{Optical} and \ref{radiodata} present the optical and 
radio data, respectively. Calculations of the energetics associated with the central AGN are discussed in 
Section \ref{centralagn}, and the results are discussed and 
summarized in Section \ref{discussion}. Throughout this paper, a \LCDM \ cosmology is assumed with 
$\Omegal=0.73$, $\Omegam=0.27$, and $H_{0}=70 \km \s^{-1} \Mpc^{-1}$. 
At the redshift of \MACS \ ($z=0.352$), 1 \arcsec \ corresponds to 4.926 \kpc.

\section{\cha \ Data Reduction and Processing}\label{datared}

Two \cha \ observations of  \MACS \ were performed using the Advanced CCD Imaging Spectrometer (ACIS) in October
2002 and August 2008. The standard level-1 event lists produced by the Chandra pipeline processing were reprocessed
using the CIAO (version 4.1.2) software package, including the appropriate gain maps and calibration products
(CALDB version 4.1.2). Bad pixels were removed and standard grade
selections were applied to the event lists. Both observations were taken in VFAINT mode, 
and the additional
information available in this mode was used to improve the rejection of cosmic ray events. The data were cleaned to
remove periods of anomalously high background using the standard energy ranges and binning methods recommended
by the \cha \ X-ray Center. The net exposure times after processing are summarized in Table \ref{Observations}.

\begin{table}
\caption{\label{Observations}Summary of the two \cha \ observations of 
\MACS. Exposure times are the net exposure after all cleaning and processing as described in Section \ref{datared}. }
\centering
\begin{tabular}{ c  c c c} 
\\ \hline {Obs \#} & {Observation Date} & Detector & Exposure Time (ks)\\ \hline\hline 

3282 & October 20 2002 & ACIS-I & 10.0 \\
9382 & August 21 2008 & ACIS-I & 89.5\\ 
\hline
\end{tabular}
\end{table}

\section{X-ray Imaging Analysis}\label{imaging}

\begin{figure}
\includegraphics[width=\columnwidth, angle=270]{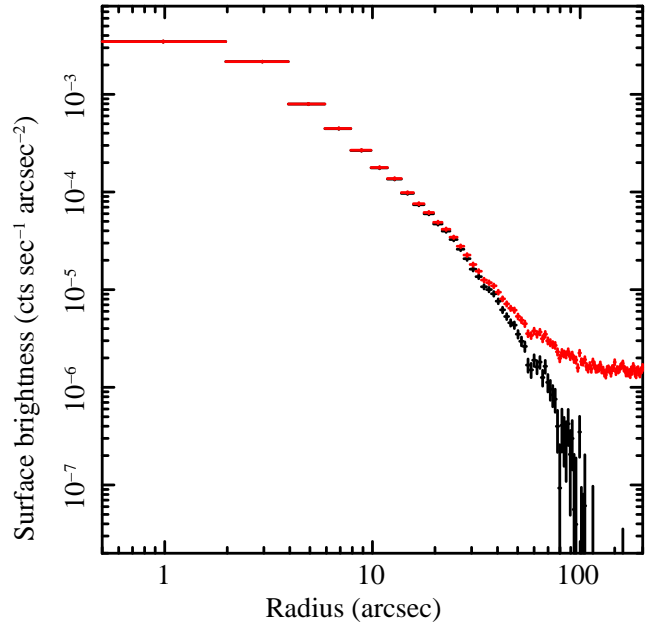}
\caption{\label{SurBrightProf} Combined X-ray surface brightness profile for \MACS \ in the energy range 0.7-2.0 \keV.
 The red points denote the total surface brightness, while the black 
points denote the background subtracted surface brightness.}
\end{figure}

\subsection{Surface Brightness Profiles}
Flat-fielded images 
were first created in the energy range from 0.7-2.0 \keV \ for each of the \cha \ observations. 
This energy range
was chosen to minimize the impact of astrophysical and instrumental background components. All subsequent imaging analysis was performed in this energy range.
Surface brightness profiles were produced from these flat-fielded
images centered on the central AGN ($\alpha (2000)=19^{h}31^{m}49^{s}.6, \ \delta (2000)=-26^{h}34^{m}33^{s}.6$). Each surface brightness profile was fit 
with a King plus constant model in the range of 50
to 400 \arcsec \ from the central AGN. The King plus constant model is explicitly given by the form  
 
\begin{equation}\label{KingModel}
I(r)=\mysub{I}{0}\left(1+\left(\frac{r}{\mysub{r}{0}}\right)^{2}\right)^{-\beta}+ \mysub{C}{0}
\end{equation}

\noindent The best-fit constant for each observation was subtracted
from each surface brightness profile, and Fig.
\ref{SurBrightProf} shows the exposure-time weighted average of the two surface brightness profiles. 
Over the radial range of 10-250 \arcsec, the profile can be approximately described by a   
 King model with $\mysub{r}{0}=6.54 \pm 0.23 \arcsec$ and $\beta=1.35 \pm 0.01$. 

The background level determined from the surface brightness analysis described above was 
subtracted from each flat-fielded image on a pixel-by-pixel basis. 
These two images were reprojected into a common aspect solution and combined. 
This combined image, which is used in all subsequent imaging 
analysis, is shown in Fig. \ref{NicerImages}.

\subsection{Background Subtracted, Flat-Fielded Image}\label{basicimaging}

\begin{figure*}
\centering
\subfigure[]{
\includegraphics[width=0.6\textwidth]{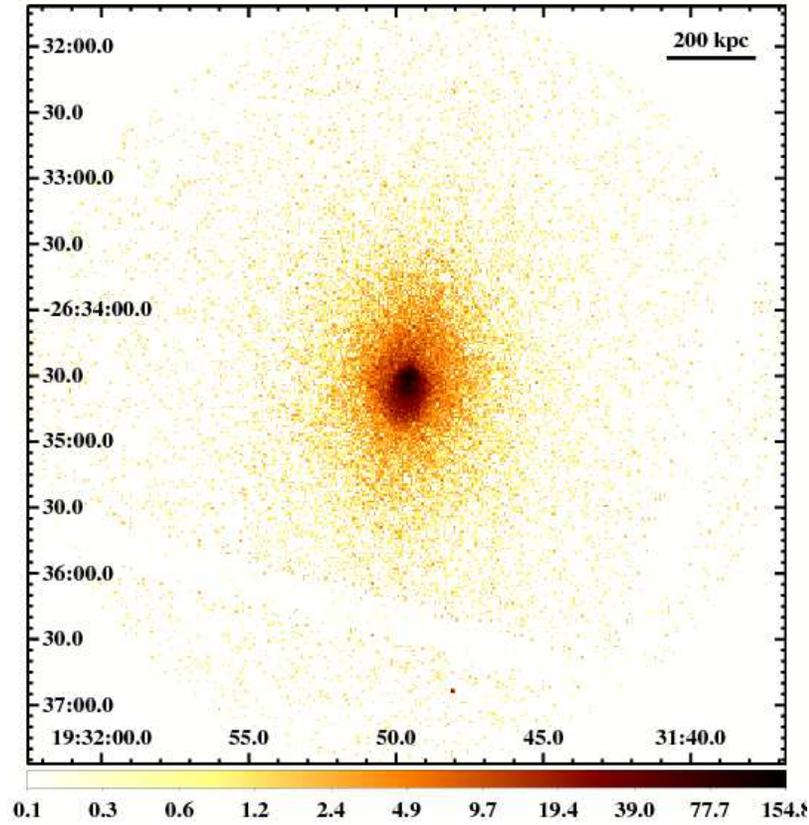}
\label{MACSBasic}
}
\subfigure[]{
\includegraphics[width=0.47\textwidth]{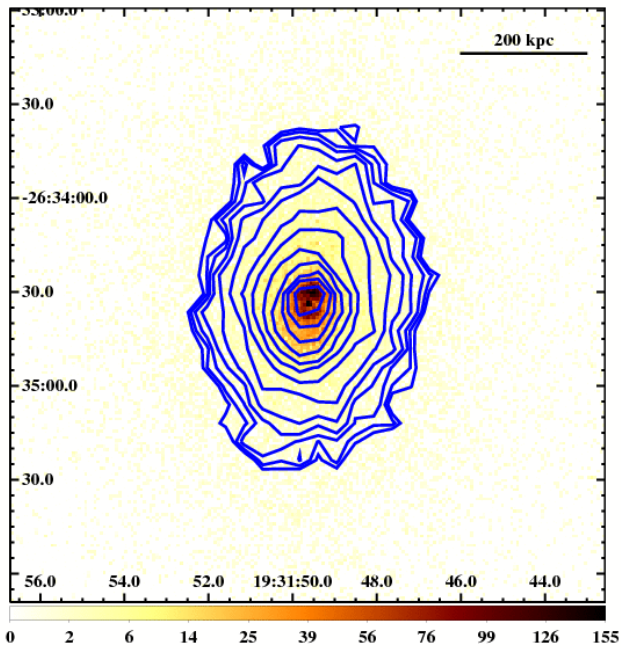}
\label{MACSCont}
}
\subfigure[]{
\includegraphics[width=0.47\textwidth]{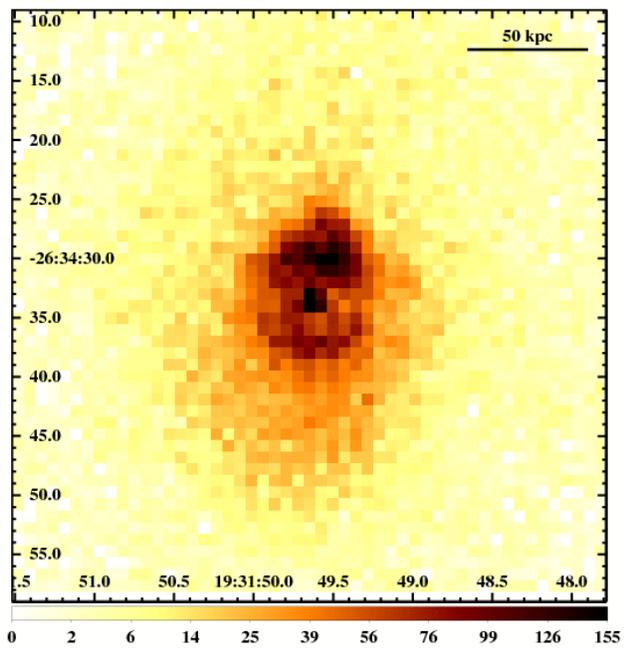}
\label{MACSZoom}
}

\caption{\label{NicerImages} Combined, background subtracted,
  flat-fielded image of \MACS \ in the energy range of 0.7-2.0 \keV. a) The central $5\arcminf 7 \times 5\arcminf 7$ of \MACS.  
b) Same as in (a), but zoomed in by a factor of about 2 and overlaid with logarithmic surface brightness contours in blue. The centroids of these
contours shift north and south of the central AGN by distances with an amplitude of up to $\sim$20-30 $\kpc$.
 c) Same as in (a) but zoomed in by a factor of about 8 to focus on the central AGN, the bright ridges about 
25 \kpc \ to the north and south, and the diffuse emission extending further to the south.}
\end{figure*}

On scales larger than $\sim$100 $\kpc$, \MACS \ exhibits a general elliptical symmetry, 
with the major axis aligned along 
the approximately north-south direction. There are asymmetries at these scales, however, that can be more readily observed
by overlaying the X-ray surface brightness contours, seen in Fig. \ref{MACSCont}. The isophotes are clearly not concentric, and the centroid appears
to shift north and south of the central AGN with an amplitude as large as $\sim$20-30 $\kpc$. 
Inside the central 100 \kpc, seen in Fig. \ref{MACSZoom}, the morphology becomes considerably more complex. 
The bright central point source is surrounded by two bright ridges to the north and south. The northern ridge is brighter and 
has a relatively sharp boundary to the north, 
while the southern ridge is trailed by diffuse emission extending further to the
south. These features, as well as the varying isophote centroids on large scales are indicative of past oscillations of the core, 
which currently appears to be moving to the north. 

\subsection{Substructure Analysis}
\subsubsection{Bandpass Filtering}
\begin{figure*}
\subfigure[]{
\includegraphics[width=0.47\textwidth]{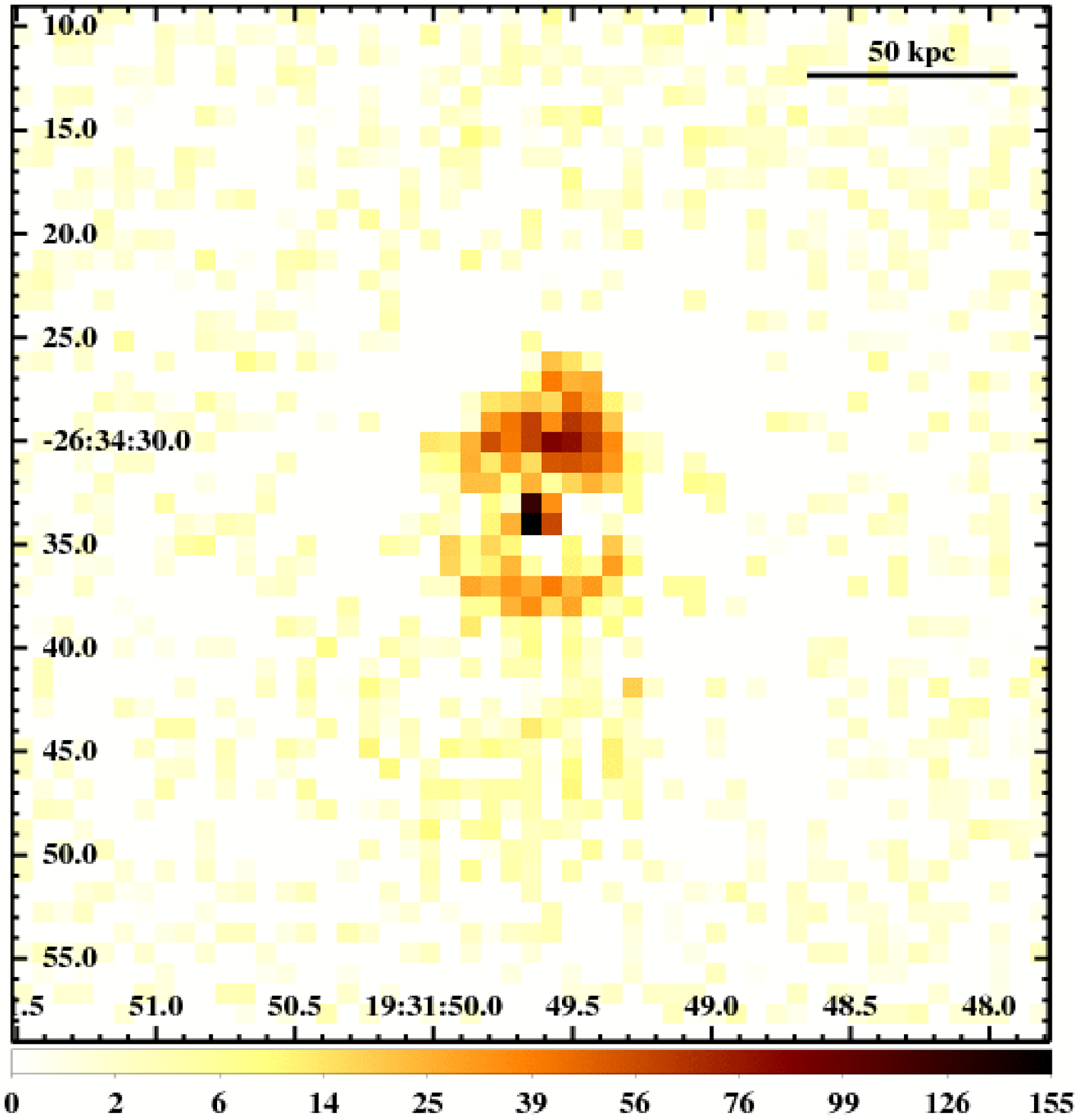}
\label{Butterworth}
}
\subfigure[]{
\includegraphics[width=0.47\textwidth]{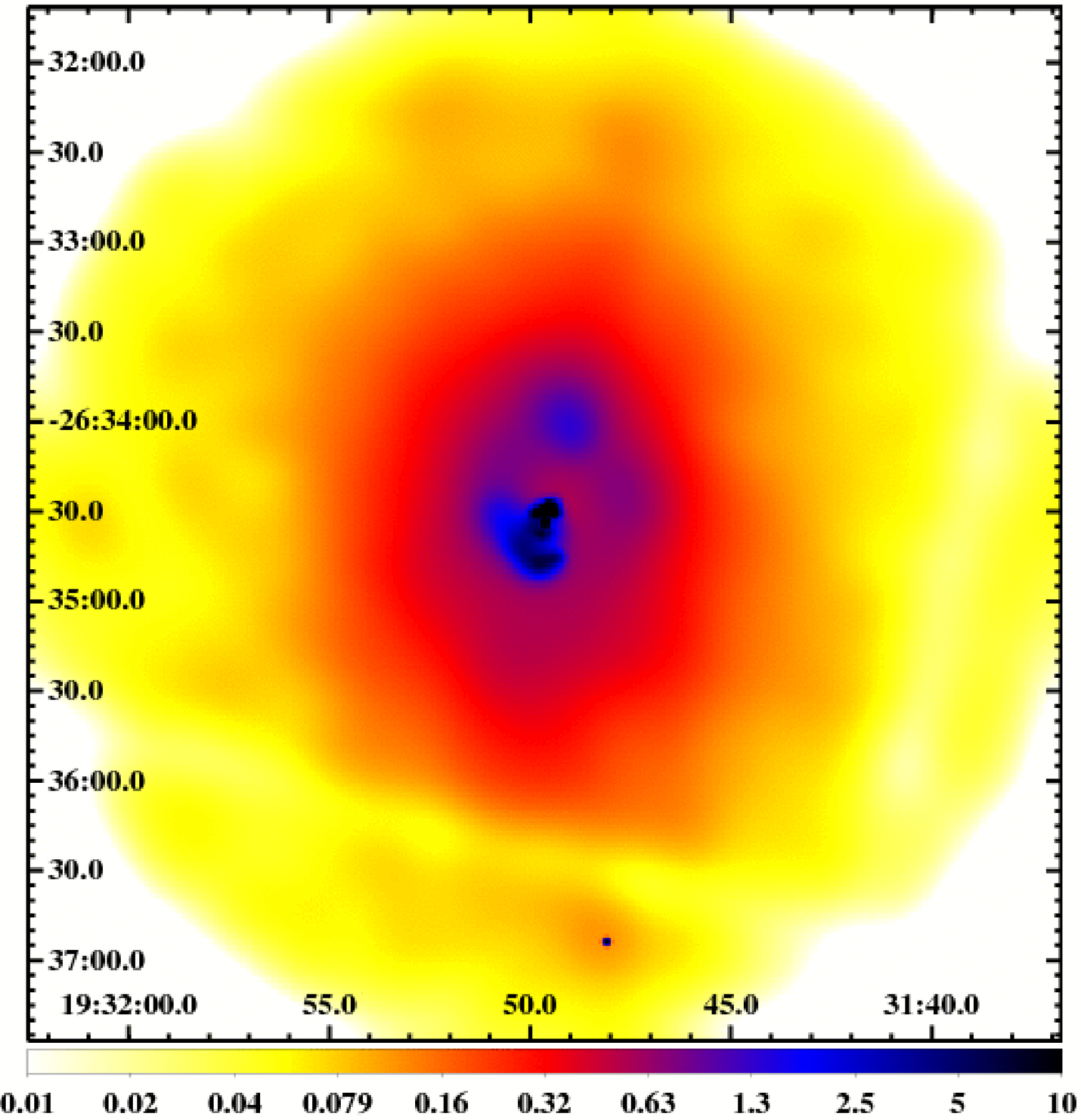}
\label{EllipBeta}
}
\caption{\label{Transformed} Images of \MACS \ that emphasize substructure.
 a) Image of \MACS \ 
after applying a high-frequency bandpass filter. This image better shows the central AGN and bright ridges to the north and south. Depressions in the X-ray emission 
also arise to the east and west of the central AGN.
b) Image of \MACS \ after subtracting the best-fit elliptical-$\beta$ continuum model and adaptively smoothing the residuals. This image better shows a large scale 
($\sim$200 $\kpc$) spiral feature wrapping around the center of the cluster.}
\end{figure*}

In order to better resolve small scale structures around the center of the cluster, 
a high-frequency bandpass filter was applied to the image
 shown in Fig. \ref{MACSBasic}. The functional form of the filter is given as  

\begin{equation}
F(k)=\frac{\left(\frac{k}{\mysub{k}{0}}\right)^{2}}{1+\left(\frac{k}{\mysub{k}{0}}\right)^{2}}
\end{equation}
\noindent We have set the scale length $\mysub{k}{0}=10$ \arcsec.
The transformed image is shown in Fig. \ref{Butterworth}, and shows the central AGN as well as the ridges to the north and south more clearly. 
Other features more apparent in this image are depressions
in the X-ray brightness immediately to the east and west of the
central AGN. Nearby systems show clear cavities in the X-ray brightness near the central AGN, similar in shape to these \citep[e.g.][]{Fabian2003,Birzan2004}.
 Unlike the cavities in those systems, however, there is no evidence in Fig. \ref{Butterworth} as to where the outer boundary of these cavities might lie.  

\subsubsection{Two-Dimensional Surface Brightness Modeling}
 The X-ray image of Fig. \ref{MACSBasic} was also fit with a two-dimensional elliptical beta model of the form 
\begin{equation}
\centering
I(x,y)= \frac{A}{\left[1+\frac{r(x,y)^{2}}{\mysub{r^{2}}{0}}\right]^{\alpha}},
\end{equation}

\noindent where $A$ is the central normalization, $\mysub{r}{0}$ is the core radius, 
and $\alpha$ is the power index.
The term $r(x,y)$ is given as 

\begin{eqnarray*}
r(x,y)^{2} &= &\left[(x-\mysub{x}{0})\cos{\theta}+(y-\mysub{y}{0})\sin{\theta}\right]^{2}\\ 
&+ &  \left[\frac{(y-\mysub{y}{0})\cos{\theta}-(x-\mysub{x}{0})\sin{\theta}}{1-\epsilon}\right]^{2},
\end{eqnarray*}

\noindent which gives the distance of any point in the image $(x,y)$ from a fixed center 
$(\mysub{x}{0},\mysub{y}{0})$ with an ellipticity $\epsilon$.
The position angle, $\theta$, was also a free parameter in the fit. 
Since the image has already been background subtracted, no further
considerations for the background were included in the fit. 
The best fit parameters for this model are $\mysub{r}{0}=3.21 \pm 0.04 \arcsec \
$,\ $\epsilon=0.290 \pm 0.004 $, $\theta=4.11 \pm 0.46 \ $ degrees and $\alpha=1.139 \pm 0.004$. 

The fit was then subtracted from the image, and the residuals adaptively smoothed. The smoothed 
residuals image is shown in Fig. \ref{EllipBeta}.
With the elliptical model subtracted, a spiral pattern beginning just east of the 
central AGN emerges. Such a pattern is expected to arise in off-axis 
mergers \citep[e.g.][see also Section \ref{discussion}]{Ascasibar2006}.

\section{Image Deprojection Analysis}\label{deprojanalysis}
An image deprojection analysis of \MACS \ was undertaken following the manner described in 
\citet{Allen2008} \citep[see also][]{Schmidt2007}. In brief, the azimuthally 
averaged X-ray surface brightness profile (centered on the central AGN) 
and the deprojected temperature profile are combined to simultaneously
determine the X-ray emitting gas mass and total mass profiles
of the cluster. 
We assume that the dark+luminous mass distribution follows the 
model of \citet{Navarro1995,Navarro1997} (hereafter NFW profile)
\begin{equation}
\rho(r)=\frac{\mysub{\rho}{c}(z)\mysub{\delta}{c}}{(r/\mysub{r}{s})(1+r/\mysub{r}{s})^{2}}
\end{equation}

\noindent where $\rho(r)$ is the mass density, $\mysub{\rho}{c}(z)=3H(z)^{2}/8\pi G$ is
 the critical density for closure at redshift $z$, $\mysub{r}{s}$ is the scale
radius, $c$ is the concentration parameter (with $c=\mysub{r}{200}/\mysub{r}{s}$) 
and $\mysub{\delta}{c}=200c^{3}/3\left[\ln{(1+c)}-c/(1+c)\right]$. The length scale \mysub{r}{\Delta} is defined such that the enclosed mean mass density is $\Delta$
 times the critical density of the Universe at the redshift of the cluster.   
For a given scale radius and concentration parameter, a model temperature  
profile can be calculated and compared with the observed deprojected temperature profile. The observed deprojected temperature profile 
was azimuthally averaged and binned into annular regions with roughly 
5,000 counts in each (see Section \ref{spectra}). 
The mass profile parameters were stepped over a range of values to determine best-fit
values and uncertainties using a $\chi^{2}$ minimization technique. This analysis assumed hydrostatic equilibrium and 
spherical symmetry, both of which are obviously not true in the vicinity of the central AGN. 
To account for these assumptions,
the data for the central 50 \kpc \ have been statistically down-weighted by adding 30\% systematic uncertainties to the measured temperatures.  

\begin{figure*}
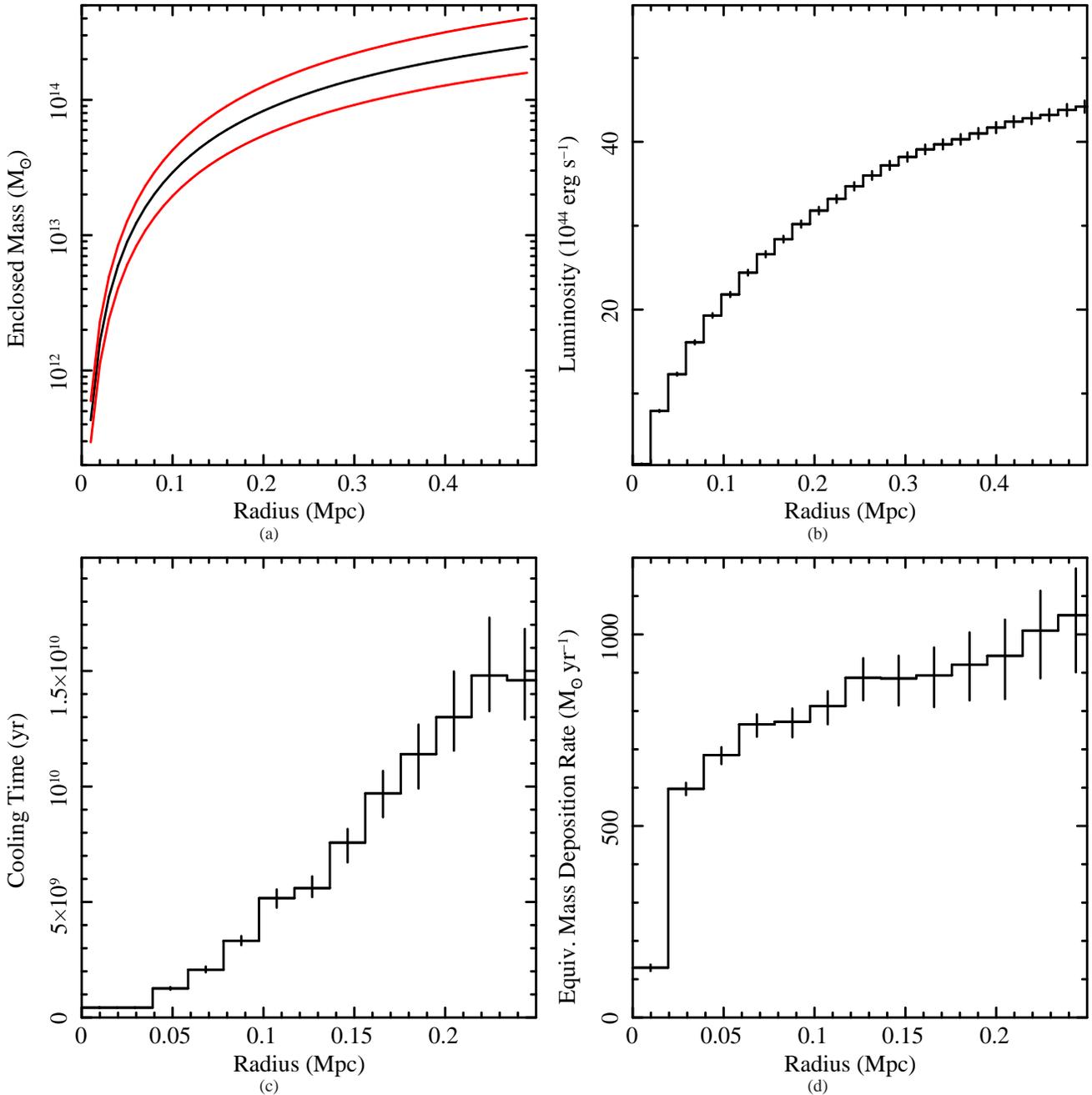

\centering
\subfigure[]{
\includegraphics[width=0.47\textwidth, angle=270]{MassProfs.ps}
\label{gravmass}
}
\subfigure[]{
\includegraphics[width=0.47\textwidth, angle=270]{Luminosity.ps}
\label{lumin}
}
\subfigure[]{
\includegraphics[width=0.47\textwidth, angle=270]{Cooltime.ps}
\label{cooltime}
}
\subfigure[]{
\includegraphics[width=0.47\textwidth, angle=270]{CoolingFlow.ps}

\label{coolflow}
}
\caption{Best fit profile from the image deprojection analysis. a) The best fit 
integrated gravitational mass profile in black, with the 
$1\sigma$ confidence interval shown as the red curves. 
 b) The integrated bolometric luminosity profile.
c) The cooling time profile. d) The equivalent mass deposition rate profile. The high luminosity within the central 50 \kpc \ leads to 
very short cooling times in this region and very large nominal mass deposition rates. For comparative purposes, the nominal mass deposition rate
in the Perseus Cluster is roughly $\sim$400 \msolaryr \citep{Fabian2002}.
}
\label{deprojfigs}
\end{figure*}

Our best fit NFW profile has a scale radius of $\mysub{r}{s}=0.26^{+0.05}_{-0.02} \Mpc$, a 
concentration parameter of $c=6.25^{+0.40}_{-0.75}$, 
and an equivalent velocity dispersion \citep{Allen2002} $\sigma=\sqrt{50}\mysub{r}{s}cH(z)=966^{+60}_{-18} \km\s^{-1}$. The integrated mass profile 
is shown in Fig. \ref{gravmass}. The
enclosed mass within $\mysub{r}{2500}=505^{+19}_{-3} \kpc$ 
is  $\mysub{M}{2500}= 2.64^{+0.31}_{-0.06} \times 10^{14} \msolar$.

The bolometric luminosity, cooling time, and equivalent mass
deposition rate profiles have also been determined using the calculations described in detail in \citet{White1997}. 
These are shown in Figs. \ref{lumin}, \ref{cooltime}, 
and \ref{coolflow} respectively.  
These profiles show that there is rapid cooling within the
central 50 \kpc. Within this region and in the absence of of balancing heat sources, a cooling flow with an equivalent
mass deposition rate of $\dot{M} \sim$700 \msolaryr would be expected.  
The bolometric luminosity within this region is $\sim$1 $\times
10^{45}$\ergs, and the cooling time is less than 1 Gyr.

\section{Spectral Analysis}\label{spectra}
\subsection{Methods}\label{spectralmethods}
The \cha \ observations of \MACS \ are sufficiently deep for high signal-to-noise spectra to be extracted from relatively small regions. 
This enables us to carry out detailed, spatially 
resolved measurements of the thermodynamic quantities of the ICM. All
spectral analysis was carried out using {\small XSPEC} \rm   \citep[][version 12.5]{Arnaud2004}. 
The backgrounds for all spectral analysis were extracted directly from
the science observations, specifically from a region roughly the same distance from the center of the 
detector as the cluster, but on the diagonally opposite chip. The background regions are devoid of point sources and cluster emission.

\subsubsection{Regions of Interest}\label{spectralbins}
The spectral structure of \MACS \ was measured in both
one-dimensional radial profiles and two-dimensional maps. In
addition, the spectra of several regions of interest identified from
the imaging analysis (specifically the central AGN and bright central northern and southern ridges) were also investigated in detail. 
To account for contamination by the central AGN, the 2.5 \arcsec \ radius region
surrounding the central AGN was excluded from the spectral profiles and maps.
Region specific response matrices and ancillary response
files were created for all spectra. 

Spatially resolved spectral maps for the cluster were extracted in regions determined by the contour 
binning algorithm of \citet{Sanders2006}, which 
creates bins of equivalent signal-to-noise, following contours in surface brightness. Regions were selected to have a signal-to-noise of 30, resulting in 
approximately $1,000$ counts per bin.

Azimuthally averaged spectral profiles were measured in
annular regions with nearly equal numbers of counts. 
Initially radial profiles were made using annular bins that 
each contain roughly 3,000 counts. Measuring the metallicity requires data with a higher signal-to-noise, so we 
also carried out a similar analysis using annular bins with roughly 10,000 counts each.
The chosen center for all annuli was the location of the central
AGN. The spectral properties of each annular region were measured both in projection and deprojection. Deprojection 
was implemented using the {\small PROJCT} \rm mixing model in {\small XSPEC}. \rm

\subsubsection{Modeling the Emission}
All spectral regions were initially modeled as a single temperature optically thin plasma using the {\small MEKAL} \rm  code 
of \citet{Kaastra1993} incorporating the
Fe-L calculations of \citet{Liedhal1995} and the photoelectric absorption models of \citet{McCammon1992}. 
We used the determinations of 
solar element abundances given by \citet{Anders1989}.
The abundances of metals ($Z$) were assumed to vary with a common ratio with respect to the Solar
values. The single-temperature plasma model has three free parameters: the temperature ($kT$), the metallicity ($Z$), 
and normalization ($K$).

In each region, the spectral analysis assumed a fixed Galactic absorption column of 
$8.3 \times 10^{20} \cm^{-2}$ \citep{Kalberla2005}, consistent with the value measured directly from the 
X-ray spectra.  
The modified Cash statistic in {\small XSPEC} \rm \citep{Cash1979,Arnaud2004} was minimized to determine the best fit 
model parameters and uncertainties. All uncertainties
given are 68\%  ($\Delta C=1$) confidence intervals, unless otherwise noted. 
In the analysis of the bright ridges to the north and south of the
AGN, the spectroscopic model included an additional {\small MKCFLOW} \rm component, appropriate for a scenario where gas is assumed to cool at constant pressure
from an upper temperature down to a lower temperature.
The upper temperature and metallicity of the {\small MKCFLOW} \rm component were tied to 
their corresponding {\small MEKAL} \rm components, 
and the low temperature was fixed to $0.1 \keV$.

\subsubsection{Thermodynamic Quantities}
Several thermodynamic quantities can be calculated directly from the
best fit {\small MEKAL} \rm model parameters. The 
electron density ($n_{e}$), pressure ($P$), and entropy ($S$) of 
the ICM are derived from the {\small MEKAL} \rm temperature ($kT$) and normalization
($K$) as
\begin{equation}\label{NormDensity}
\mysub{n}{e}^{2}=\frac{4\pi \times 10^{14}\left(1+z\right)^{2}\mysub{D}{A}^{2}K}{1.2V}
\end{equation}

\begin{equation}
P=kT\mysub{n}{e}
\end{equation}

\begin{equation}
S=kT\mysub{n}{e}^{-2/3}
\end{equation}

\noindent where the cosmological value of the angular diameter distance $\mysub{D}{A}$ is 1016 \Mpc \ 
at the cluster redshift. The volume of the region, $V$, is given in units of $\cm^{3}$. If the region had not been 
deprojected (in the case of the thermodynamic maps), the volume of the region was estimated as

\begin{equation}
V=\mysub{D}{A}^{3}\Omega \sqrt{\mysub{\theta}{max}^{2}-\mysub{\theta}{min}^{2}}
\end{equation}
where $\mysub{\theta}{max,min}$ are the maximum and minimum angular distances of any point in the region 
 to the center of the cluster, respectively, and $\Omega$ is the solid-angle 
extent of the region in the sky \citep{Henry2004,Mahdavi2005}.
 If the spectrum had been fully deprojected (true for the annular profiles), the volume was calculated as 
\begin{equation}
V=\frac{4}{3}\pi\mysub{D}{A}^{3}\left(\mysub{\theta}{max}^3-\mysub{\theta}{min}^{3}\right)
\end{equation}
Since the uncertainties on the temperature are considerably larger than those on the density ($> 10\% $ for the 
temperature as compared to $\sim$2-5\%  for the densities), the fractional uncertainties on the pressure 
and entropy given are similar to the corresponding 
temperature measurement.

\subsection{Spectral Results}\label{thermodyn}

\subsubsection{Thermodynamic Mapping}\label{thermomapsec}
\begin{figure*}
\centering
\subfigure[]{
\includegraphics[width=0.47\textwidth]{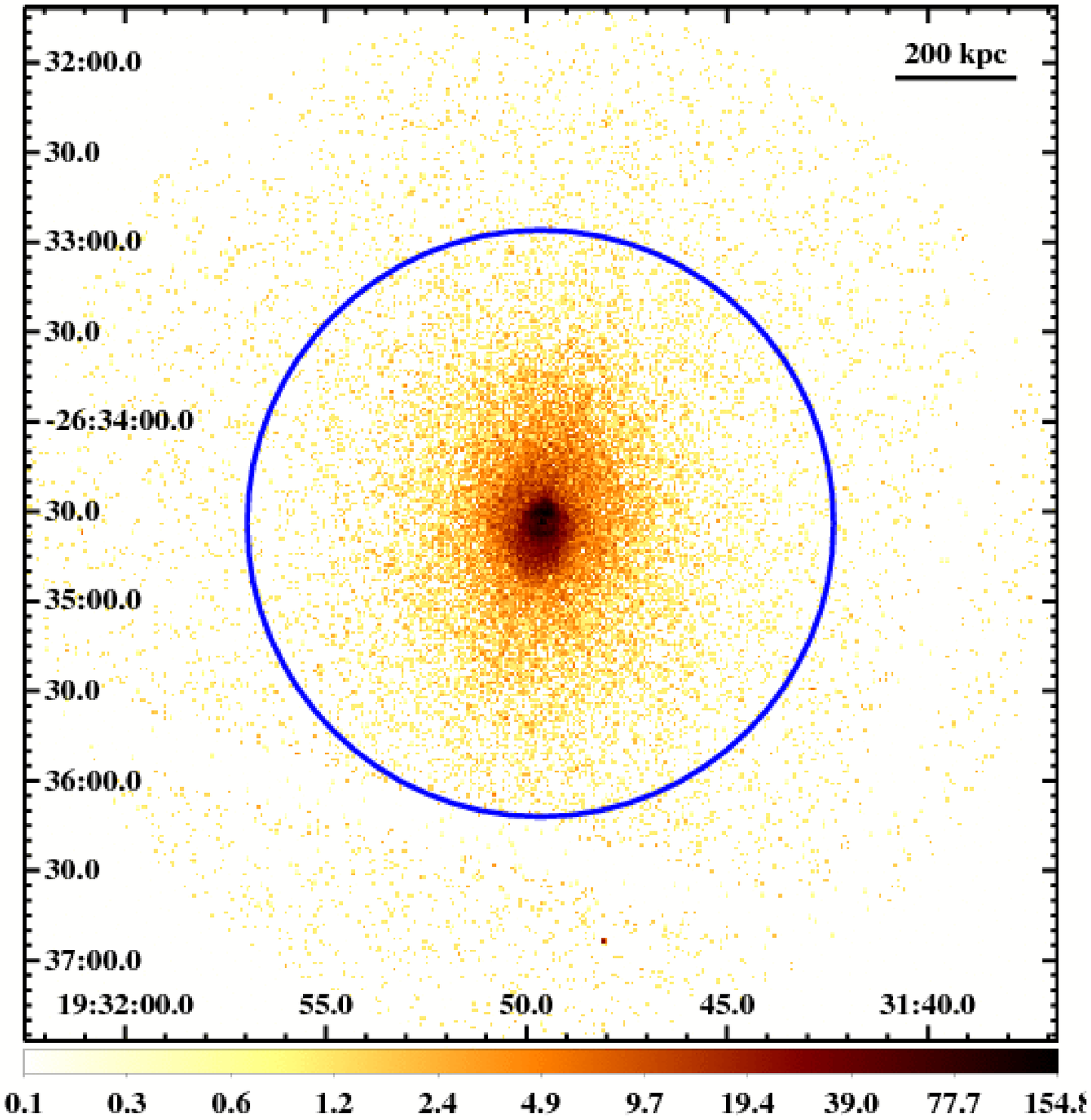}
\label{MapReg}
}
\subfigure[]{
\includegraphics[width=0.47\textwidth]{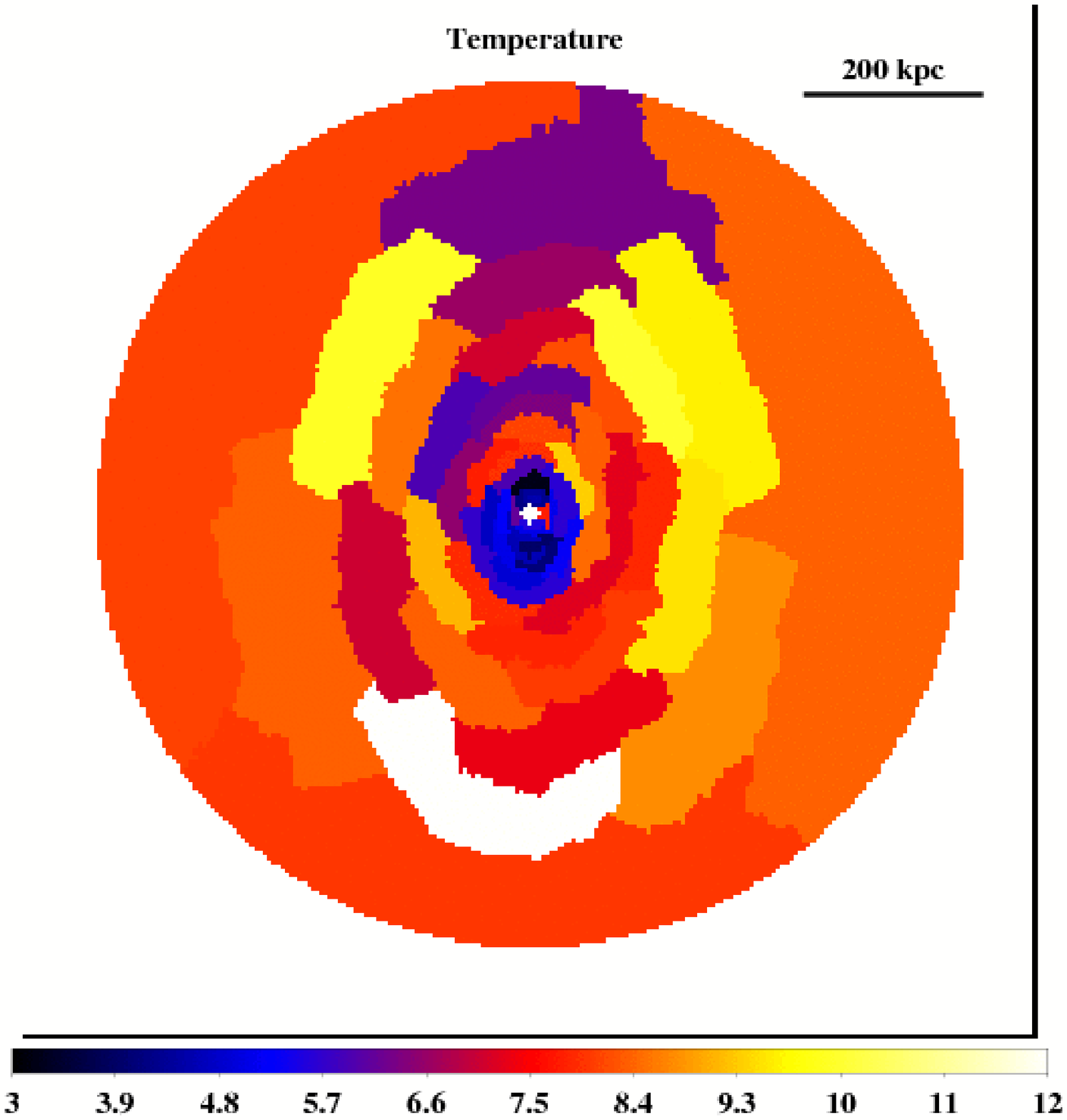}
\label{MapTemp}
}
\subfigure[]{
\includegraphics[width=0.47\textwidth]{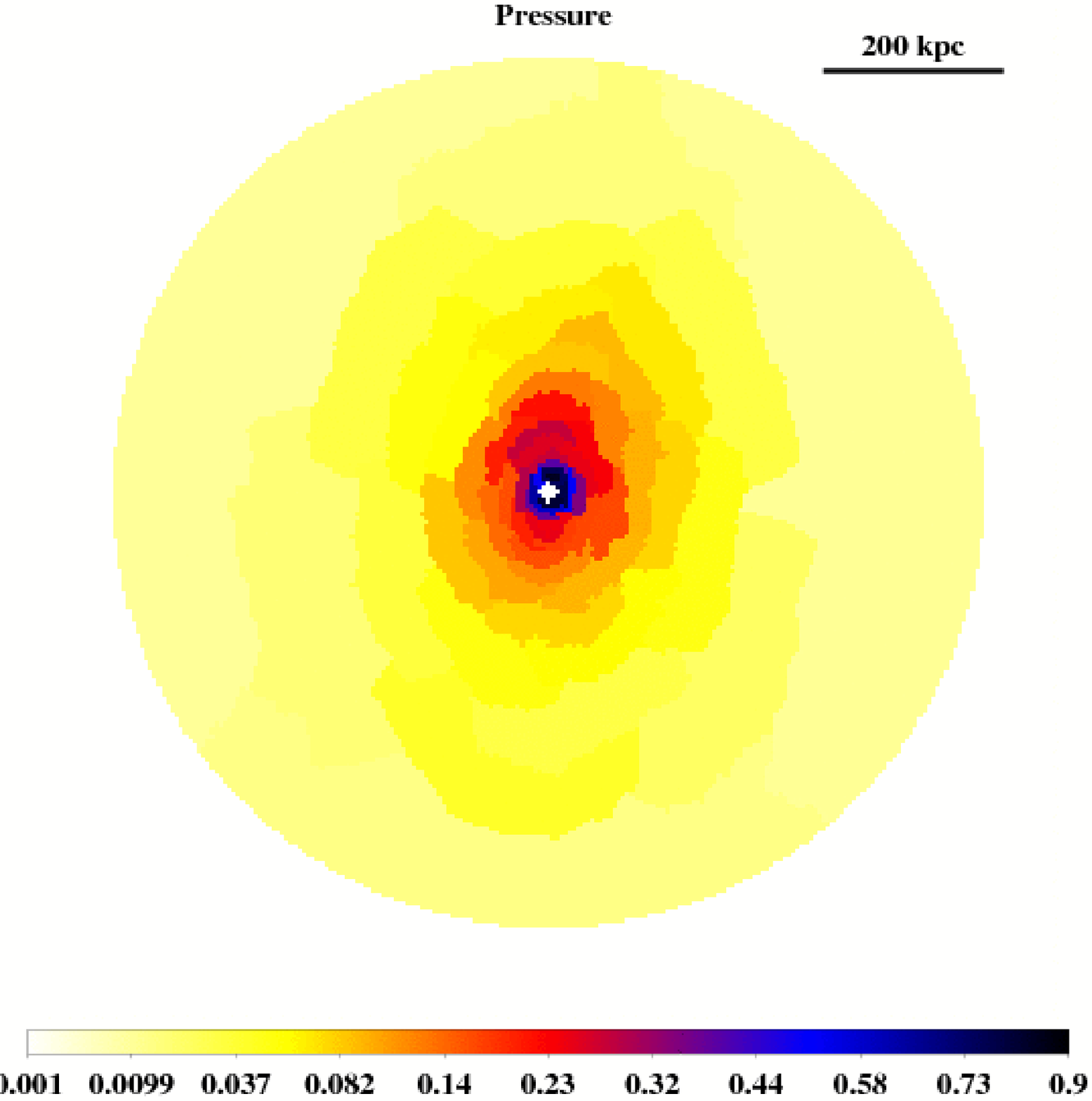}
\label{MapPressure}
}
\subfigure[]{
\includegraphics[width=0.47\textwidth]{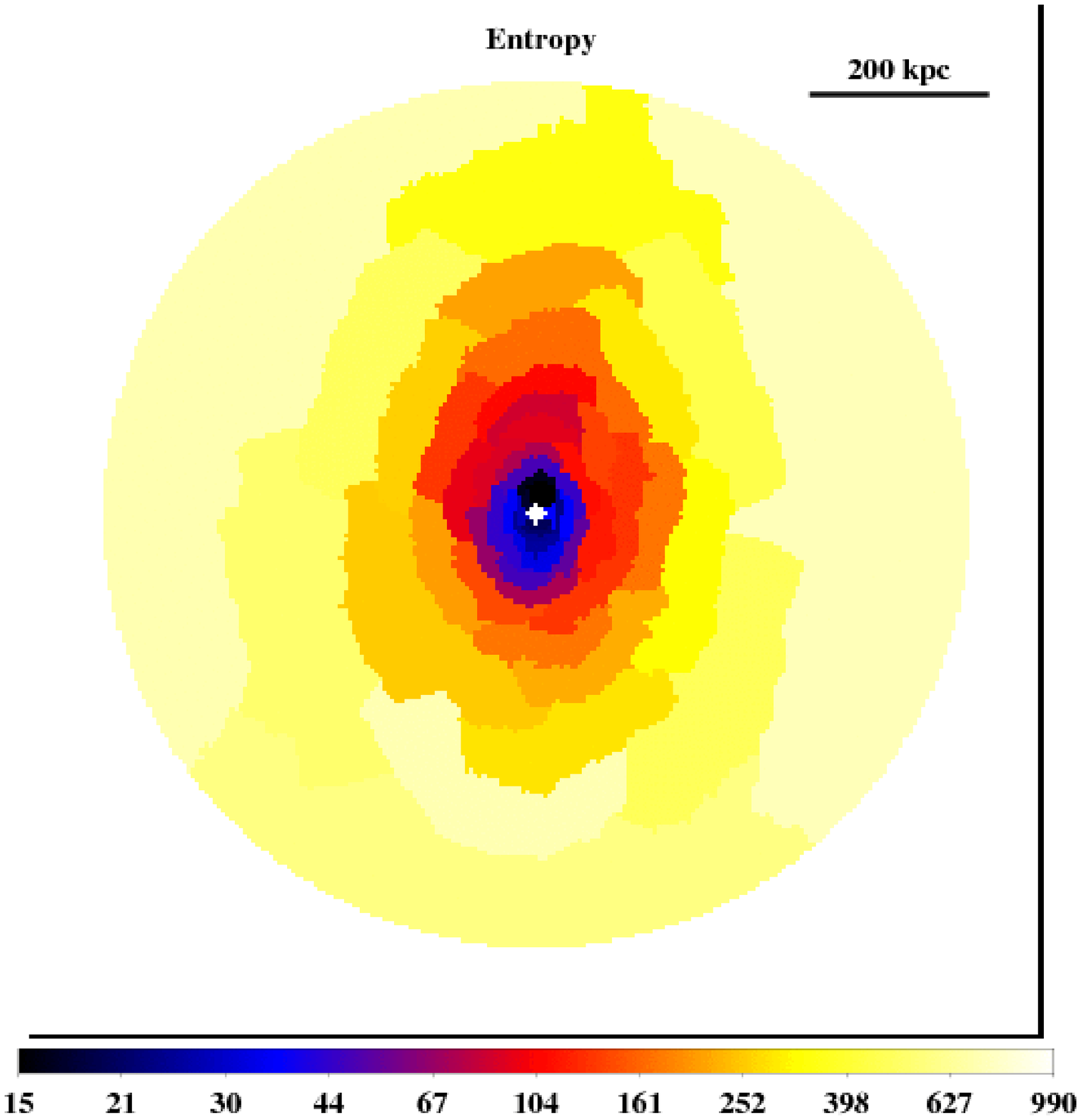}

\label{MapEntropy}
}
\caption{Thermodynamic maps for \MACS.  
a) The surface brightness image of Fig. \ref{MACSBasic} with the field of view of the thermodynamic maps overlaid in blue.
b) Temperature, $kT$, in units of \keV. 
c) Pressure, in units of keV
cm$^{-3}$.
d) Entropy, in units of keV cm$^{2}$. The 1$\sigma$ fractional uncertainties in the mapped
quantities are $\approxlt 20$ per cent. The white region at the center 
of these maps is the exclusion region for the central AGN. 
There are a total of 60 independent regions shown in these maps, which have a field of view approximately 200 \arcsec \ (1 Mpc) in diameter. 
}
\label{ThermoMaps}
\end{figure*}

The maps of temperature, density, pressure, and entropy in \MACS \ are shown in Fig. \ref{ThermoMaps}.
One of the most notable features is the spiral of low temperature 
gas wrapping to the east and north of the central AGN seen in the large scale
temperature map. This is spatially 
coincident with the surface brightness excess seen in Fig. \ref{EllipBeta}. 
Such spirals can be due to either ram pressure stripping 
of an infalling subcluster core, or merger induced oscillatory motion of the cluster core. 
Both scenarios are seen in many nearby systems and hydrodynamic simulations 
\citep[e.g.][]{Churazov2003,Ascasibar2006,Dupke2007,Lagana2009,Owers2009,ZuHone2009,Million2010}. 

\begin{figure*}
\centering
\subfigure[]{
\includegraphics[width=0.47\textwidth]{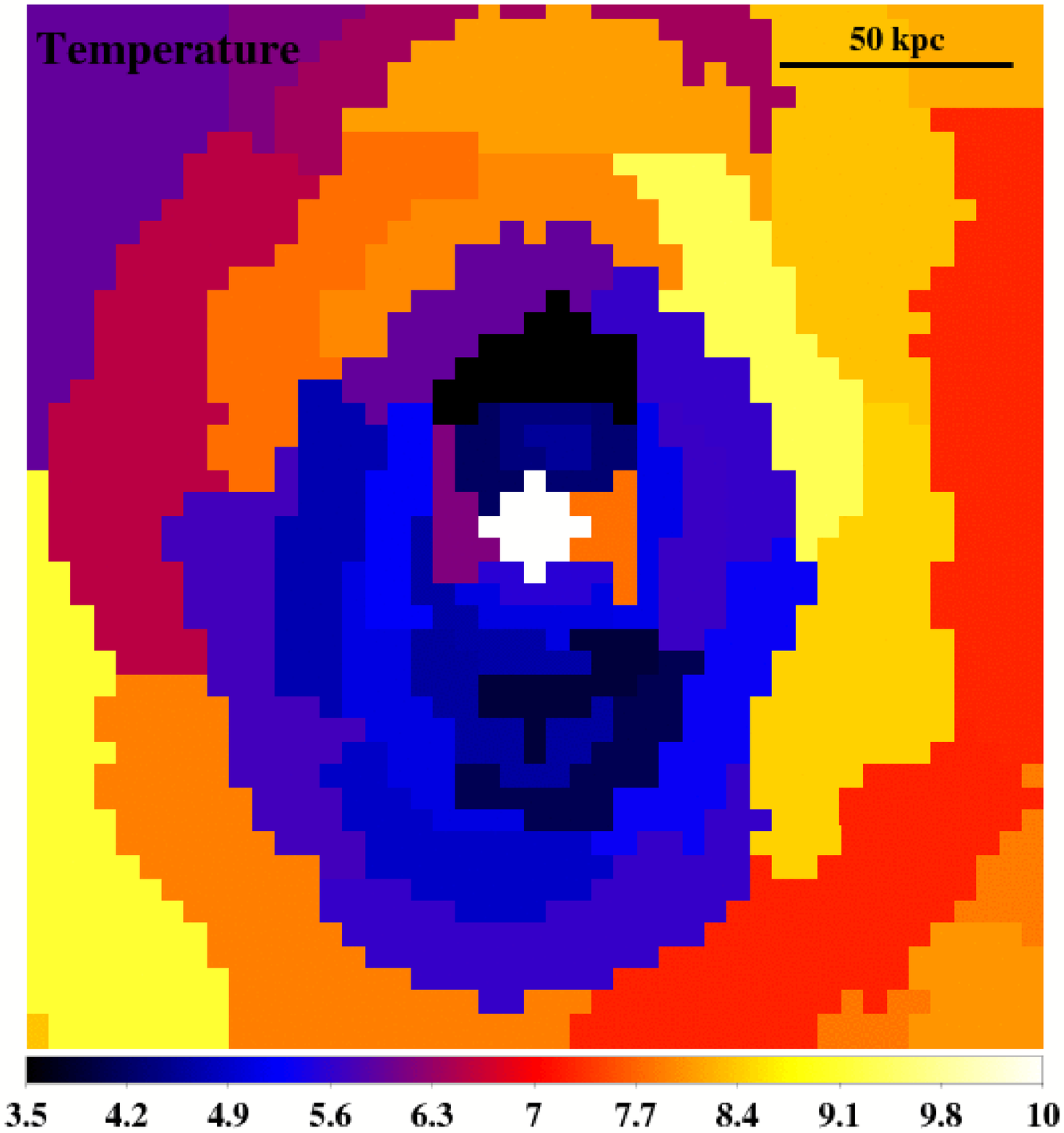}
\label{ZoomTemp}
}
\subfigure[]{
\includegraphics[width=0.47\textwidth]{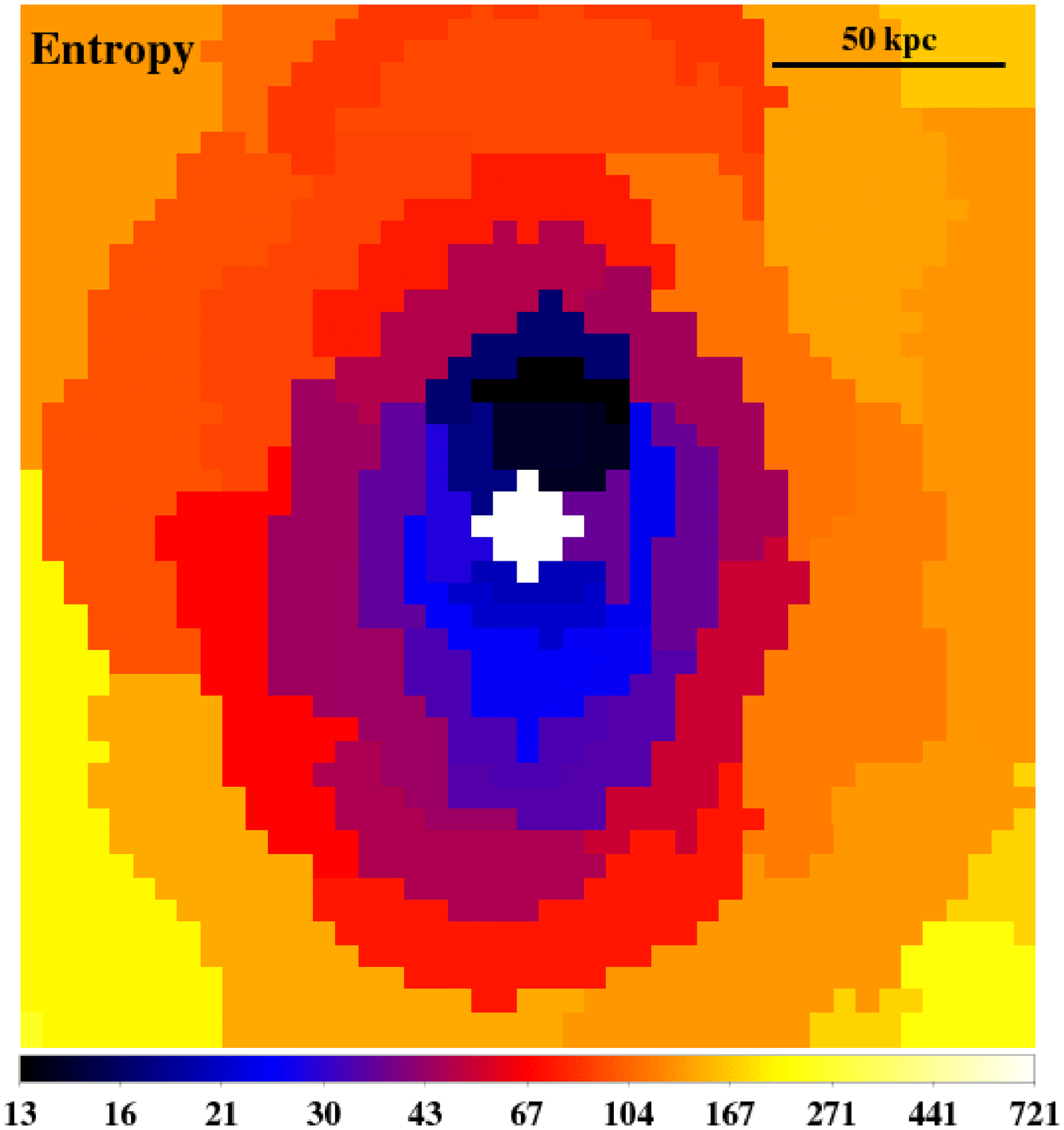}
\label{ZoomEntropy}
}

\caption{The thermodynamic structure around the central AGN. These maps are identical to those shown in Fig. \ref{ThermoMaps}, 
but zoomed in on the central AGN by approximately a factor of 4. 
a) Temperature, in keV. b) Entropy, in units of keV cm$^{2}$. It is clear in both of these maps that the lowest 
temperature and entropy gas is located approximately 25 \kpc \ to the north of the central AGN.
}
\label{ZoomedThermoMaps}
\end{figure*}

Zooming in on the temperature and entropy structure surrounding the AGN, Figs. \ref{ZoomTemp} and \ref{ZoomEntropy} show that 
there are also two small regions of higher temperature
to the east and west of the central AGN spatially coincident with 
the depressions in X-ray emission noted in Fig. \ref{Butterworth}.
Although the temperature of these regions is up to 2 \keV \ higher
than their surroundings, the statistical significance is modest. The lowest temperature and entropy gas is located 
 25-30 \kpc \ to the north of the AGN and spatially coincident with the X-ray bright northern ridge. 
There is a similar ridge to the south which also has lower temperature and entropy gas, but at lower significance.

\subsubsection{Spectroscopic Cooling in the Bright Northern and Southern Ridges}\label{ridges}

Since cooling
flows are only expected in the inner, densest regions of the ICM, we have
searched for the presence of spectroscopic cooling in the bright ridges
to the north and south of the central AGN. The regions defined as the
northern and southern ridges are shown in Fig. \ref{RidgeRegions}. These
regions were constructed to surround the brightest emission in the image shown in Fig. \ref{MACSZoom}. 

The two ridges are approximately equal distances from the central AGN, but have different spectral properties.
The northern ridge has an emission weighted temperature 
of $kT=4.78 \pm 0.64$ and a metallicity $Z=0.53 \pm 0.11\zsolar$. This metallicity is higher than the average of this cluster of $Z=0.36 \pm 0.03 \zsolar$, 
 discussed in more detail in Section \ref{metalprofsec}. Adding a cooling flow component to the spectral model of the northern ridge, 
we measure a spectroscopic mass deposition rate of $\dot{M}=165 \pm 56$ \msolaryr. 
Comparing the luminosity of the composite {\small MEKAL+MKCFLOW}
\rm model with the {\small MKCFLOW} \rm component alone, the spectroscopic cooling of 
gas down to $kT \sim$0.1 $\keV$ \  contributes $\sim$30\% of the emission from the 
northern ridge region. The southern bright ridge has a similar overall temperature (emission weighted  $kT=5.87^{+1.30}_{-0.46}$), 
but a lower metallicity ($Z=0.22 \pm 0.09$) and a 90\% confidence upper limit on the spectral 
cooling rate of $\dot{M} <83$ \msolaryr. These results are summarized in Table \ref{speccoolingtable}.

We note that the {\small MEKAL+MKCFLOW} \rm model offered a significant improvement to the spectral fit of the 
northern ridge ($\Delta C \sim$10 with one additional fit parameter, 
a significant improvement at the $\sim$99.9\% confidence level), 
almost identical to the $\Delta C$ improvement obtained with a two temperature ({\small MEKAL+MEKAL} \rm) fit, which introduces two additional fit parameters to the 
default {\small MEKAL} \rm model.
We also attempted to further resolve the spectroscopic cooling flow by using 2 or more {\small MKCFLOW} components with different temperature ranges, 
but none of these more complicated models provided a further statistically significant improvement to the fit. 
For the southern ridge, the fit was not improved by the addition of either a cooling flow or a second {\small MEKAL} \rm component.

\begin{table}
\caption{\label{speccoolingtable} Spectroscopic measurements of the X-ray bright 
ridges to the north and south of the central AGN. Temperatures are given in \keV, 
metallicities are in solar units, and the cooling flow rate is given in\msolaryr. 
Upper limits are given at the 90\% confidence level.}
 
\begin{tabular}{ c c c } 
\\ \hline Measurement & North &  South\\ \hline\hline 

$kT$ & $4.78 \pm 0.64 $ &   $5.87 ^{+1.30}_{-0.46}$   \\
\\
$Z$ & $0.53 \pm 0.11$  & $ 0.22 \pm 0.09 $\\ 
\\
$\dot{M}$ & $165^{+45}_{-67} $ &  $ <83$\\  
\\
\hline
\end{tabular}
\end{table}

\subsubsection{Azimuthally Averaged Thermodynamic Profiles}\label{thermoprofsec}

\begin{figure}
\includegraphics[width=0.47\textwidth, angle=270]{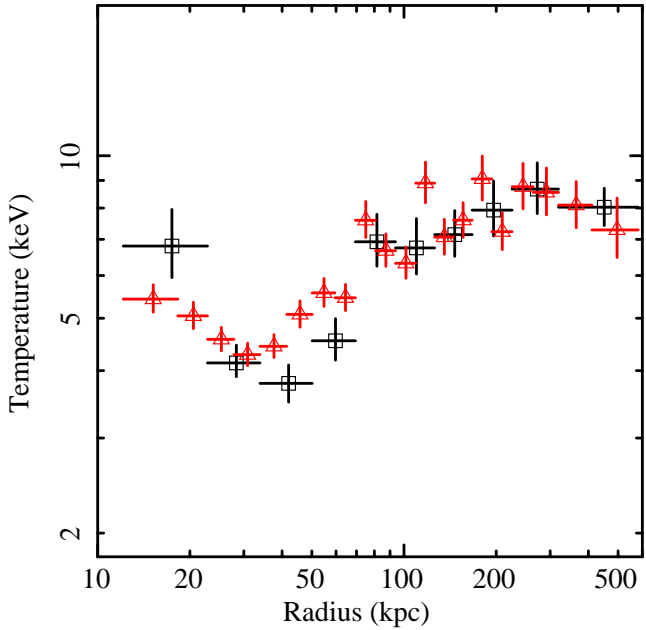}
\caption{\label{TempWRidges} The azimuthally averaged temperature profile of \MACS. The red triangles denote the projected temperature
profile, while the black squares denote the deprojected profile. The central region has a much higher temperature than the adjacent regions,
which suggests that shock heating might be present. The radius of the temperature minimum corresponds to the distance of the northern and southern ridges.}
\end{figure}

The azimuthally averaged temperature profile for \MACS \ is shown in Figure \ref{TempWRidges}. It is clear in this profile that the temperature 
does not decrease monotonically towards the
center as is expected in a cool core cluster \citep[e.g.][]{Allen2001,Vikhlinin2005}, but instead appears to increase in the
central-most regions. This central jump in temperature is most clear in deprojection, where the innermost temperature is several \keV \ higher than
the measured temperature in the adjacent region. Such a sharp discontinuity in the temperature is often associated with shock heating, but the presence of the cooler
ridges separated from the central AGN makes presence of shock heating ambiguous. 
To account for these cool ridge substructures, thermodynamic profiles were taken with the ridge regions excluded. These thermodynamic profiles are shown in Fig. 
\ref{ThermoProfs}. These profiles more accurately describe the average thermodynamic structure of the ICM. 
After excluding the ridges, the deprojected temperature in the central-most region  still does not decrease towards the center. The sharp temperature jump seen 
in the center of the deprojected temperature profile is less pronounced, but still present, and there are weak indications of discontinuities in the central bin of the 
density profile as well. All of the thermodynamic profiles of Figure \ref{ThermoProfs} are consistent with shock heating occurring in the central 20 \kpc, but 
the small number of regions and relatively low signal-to-noise do not allow for us to claim an unambiguous detection. Such a jump in the central temperature could also
be due to the presence of cavities devoid of ICM gas. In that case, the measured central temperature would be due to ICM gas in front of and behind the central AGN. 
This should lead to a flatter deprojected density profile in the central regions, a feature that is not observed. 
The innermost annular region of the thermodynamic profiles is also considerably larger than any apparent cavities in the X-ray emission, 
so it is unlikely that this temperature jump is due to projection effects involving cavities.  
 
Both the temperature and density profiles show other discontinuities within the central 100 \kpc.  The density profile decreases discontinuously at several locations
while the temperature profile increases sharply at $r \sim$70 $\kpc$. These profiles lead to clear discontinuities in the entropy and 
an unusually flat pressure profile. Such discontinuities arise from the presence of either cold fronts or weak ($\mach < 2$) shocks. 
Since the lower temperature gas on the inner edges of these fronts has a higher density, this profile is more consistent with the 
presence of cold fronts than shock heating. Heating from weak shocks cannot be ruled out, however, since the state of the ICM at earlier times before 
any bulk motion or shock heating is unknown. Deeper data are required to distinguish between these possibilities.

\subsubsection{Metallicity Profiles}\label{metalprofsec}
\begin{figure}
\includegraphics[width=\columnwidth, angle=270]{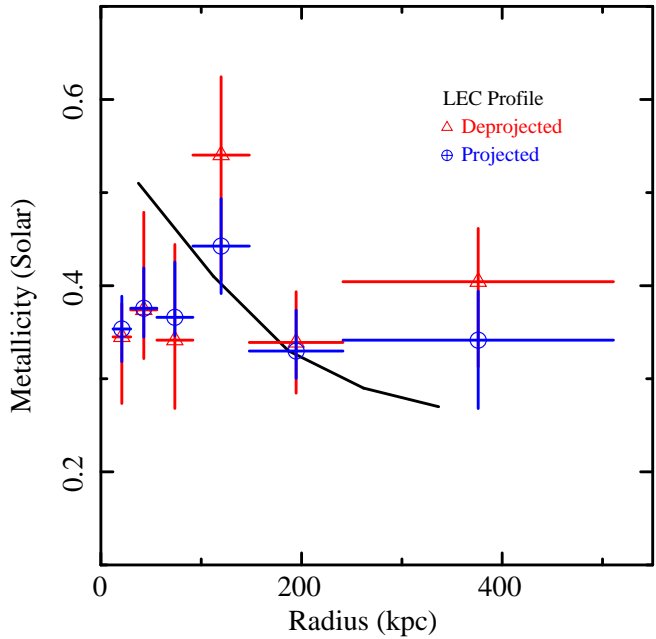}
\caption{\label{Metals} Azimuthally averaged metallicity profiles of \MACS. The black curve is the mean 
profile derived for the Low Entropy Core (LEC)
sample from \citet{Leccardi2010} scaled to the estimated $\mysub{r}{180}$ for \MACS. The blue circles are the projected metallicity profile, 
while the red triangles are the deprojected metallicity profile. 
}
\end{figure}

Surprisingly, the metallicity profile of \MACS \  
shown in Fig. \ref{Metals} exhibits no deviations from a constant metallicity of $Z=0.36\zsolar$ out to distances as large as 400 \kpc.
 All attempts to fit the data to a multi-temperature {\small MEKAL} \rm model did not
lead to any significant increases in the central metallicity. 
A central metallicity peak is almost always observed in the profiles for cool core clusters \citep{Allen1998,DeGrandi2001,Leccardi2008,Leccardi2010,Ehlert2009}. 
The only region in the cluster that has a significant metallicity
enhancement is the bright ridge of X-ray gas to the north of the central AGN, where the metallicity is closer to $Z=0.5\zsolar$. 
The unusual metallicity profile of \MACS \ argues 
that the cool core may have undergone substantial 
stripping, likely associated with its bulk motion. 
A dramatic example of stripping of a cool core due to bulk motion has
been recently reported in the nearby Ophiuchus Cluster \citep{Million2010}.

\begin{figure}
\centering
\includegraphics[width=0.47\textwidth,height=0.53\textwidth ]{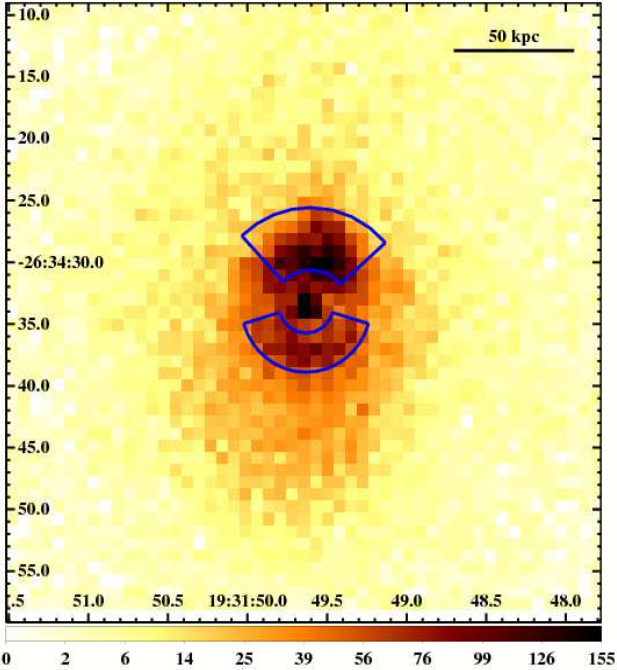}
\caption{The regions defined as the bright northern and southern ridges, drawn over the image of Fig. \ref{MACSZoom} in blue. These encapsulate the 
regions of the brightest X-ray emission in the cluster apart from the central AGN. The spectral properties of these regions are discussed in detail in Section
\ref{ridges}.  }
\label{RidgeRegions}
\end{figure}

\section{Optical structure of the cluster core}\label{Optical}

\begin{figure*}
\begin{minipage}{0.57\hsize}
\includegraphics[width=\hsize]{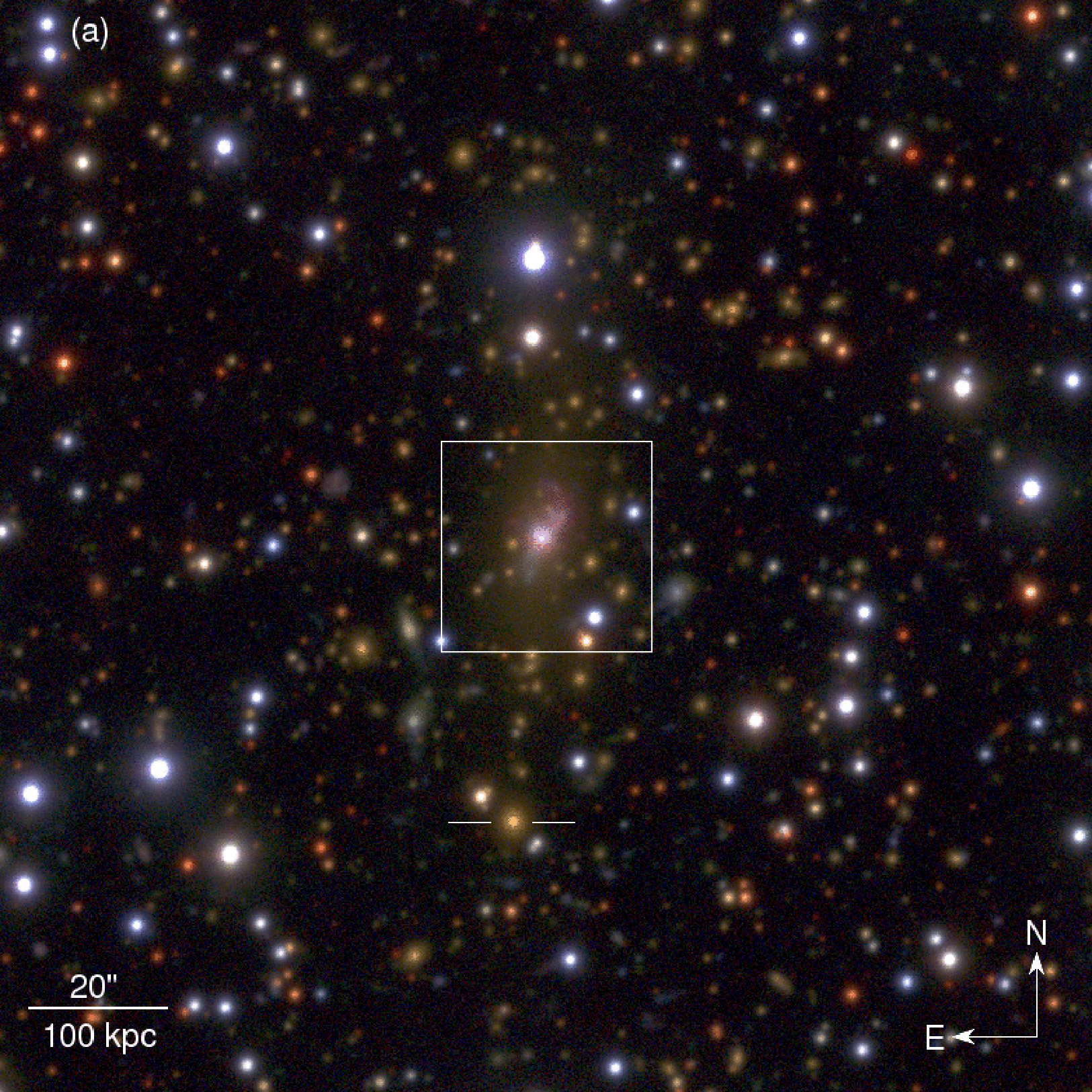}
\end{minipage}%
\hspace{0.01\hsize}%
\begin{minipage}{0.41\hsize}
\includegraphics[width=\hsize]{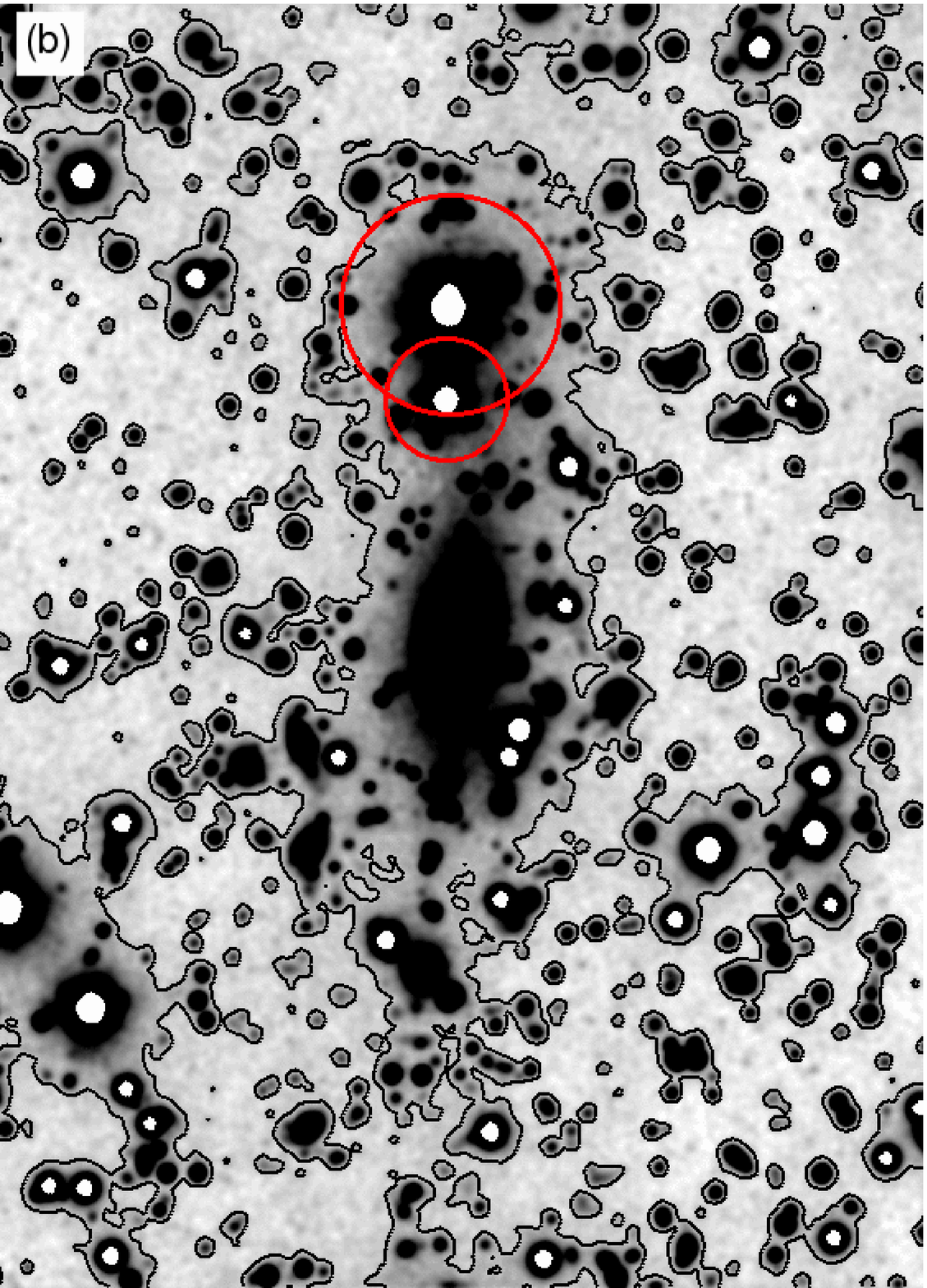}
\end{minipage}%
\caption{(a): {\it BRz} image of the central $2\arcminf6 \times
  2\arcminf6$ ($780{\rm kpc} \times 780{\rm kpc}$) of
  MACS~J1931.8-2634. The sightline to MACS~J1931.8-2634 is close to
  the galactic center, and thus most objects in this image are
  foreground stars. The box indicates the field of
  Fig.~\ref{fig:bcg_zooms}. (b): The deepest band, {\it R}, smoothed
  with a $0\arcsecf6$ Gaussian kernel, and scaled to bring out
  low-surface brightness features. The black contour follows a surface
  brightness level of 28.3~mag arcsec$^{-2}$. Bright stars are marked
  as white circles. For the two stars north of the BCG, red circles
  indicate where the PSF surface brightness reaches $\sim$28.3~mag
  arcsec$^{-2}$.  Note how the BCG envelope and the intra-cluster
  light are highly elongated in the N-S direction.}
\label{fig:core_optical}
\end{figure*}

We observed MACS~J1931.8-2634 in broad-band {\it BVRIz} filters with
SuprimeCam on Subaru (Table~\ref{tab:optical}). The data were reduced
with a dedicated weak lensing and photometry pipeline based on the
GABODS pipeline of \citet{esd05} as part of a larger cluster sample
(von der Linden et al., in
preparation). Figure~\ref{fig:core_optical}(a) shows a {\it BRz}
three-color image of the central $2\arcminf6 \times 2\arcminf6$ of the
cluster.

\begin{table}
  \caption{Summary of the optical data. All images were taken with SuprimeCam at the Subaru telescope. The quoted seeing values are those of the coadded images.}
\label{tab:optical}
\begin{tabular}{l c c c}
Filter & Exposure time [s] & Observation date & Seeing\\
\hline
{\it B} & 1680 & 2006-06-25 & $0\arcsecf89$\\
{\it V} & 1636 & 2006-05-30 & $0\arcsecf83$ \\
{\it R} & 3360 & 2006-06-25 & $0\arcsecf76$ \\
{\it I} & 2400 & 2006-05-30 & $0\arcsecf88$ \\
{\it z} & 1620 & 2007-07-18 & $0\arcsecf71$ \\
\end{tabular}
\end{table}

Similar to the X-ray structure, the optical morphology of the cluster
core exhibits clear structure in the North-South direction: the
Brightest Cluster Galaxy (BCG) and the intracluster light (ICL) are
highly elongated in this direction.  In Fig.~\ref{fig:core_optical}(b)
we highlight the extent of the ICL. The SuprimeCam images are very
deep, and we have taken significant care in flat-fielding the images
(von der Linden et al., in preparation). This allows us to trace the
ICL to $\sim$28.3~mag arcsec$^{-2}$. Correcting for cosmological
surface brightness dimming, this would correspond to 27~mag
arcsec$^{-2}$ at $z=0$. North of the BCG, two stars mask the ICL, but
there is evidence that the ICL extends slightly northwest of the
stellar halos. To the south, the ICL appears to encompass the second
brightest galaxy, located $\gtrsim 200$~kpc south of the BCG. At these
extremes, the detected ICL extends to $\sim 200$~kpc north and south
of the BCG, but only to $\sim 70$~kpc east and west. The ICL of
MACS~J1931.8-2634 is thus highly elongated, with an axis ratio of
$\sim 0.3$. At radii $\gtrsim 50$~kpc, the ICL traces the overall
gravitational potential and is tied to the evolution history of the
cluster \citep[e.g.][]{kzw02,npa03,gzz05}.  The stars in the ICL are
collisionless, like the dark matter, and thus the shape of the ICL
should presumably reflect the shape of the core of the cluster dark matter
halo. The elongation of the dark matter distribution indicates the
direction of the last merger \citep[e.g.][]{rlb97}, and thus
MACS~J1931.8-2634 likely experienced a merger within the N-S
direction. The timescale of this event is not clear, since the
elongation persists for several Gyr \citep{rlb97,mkd04}. A
confirmation of this elongation of the dark matter halo by
gravitational lensing is currently not possible - the ground-based
data do not reveal any strong lensing features. Furthermore, the
sightline to MACS~J1931.8-2634 is close to the galactic center
($l=12.5669\arcdeg$, $b=-20.09\arcdeg$), and the field is crowded with
stars, which prohibit weak lensing shape measurements for an
appreciable number of background galaxies.

\begin{figure*}
\includegraphics[width=0.33\hsize]{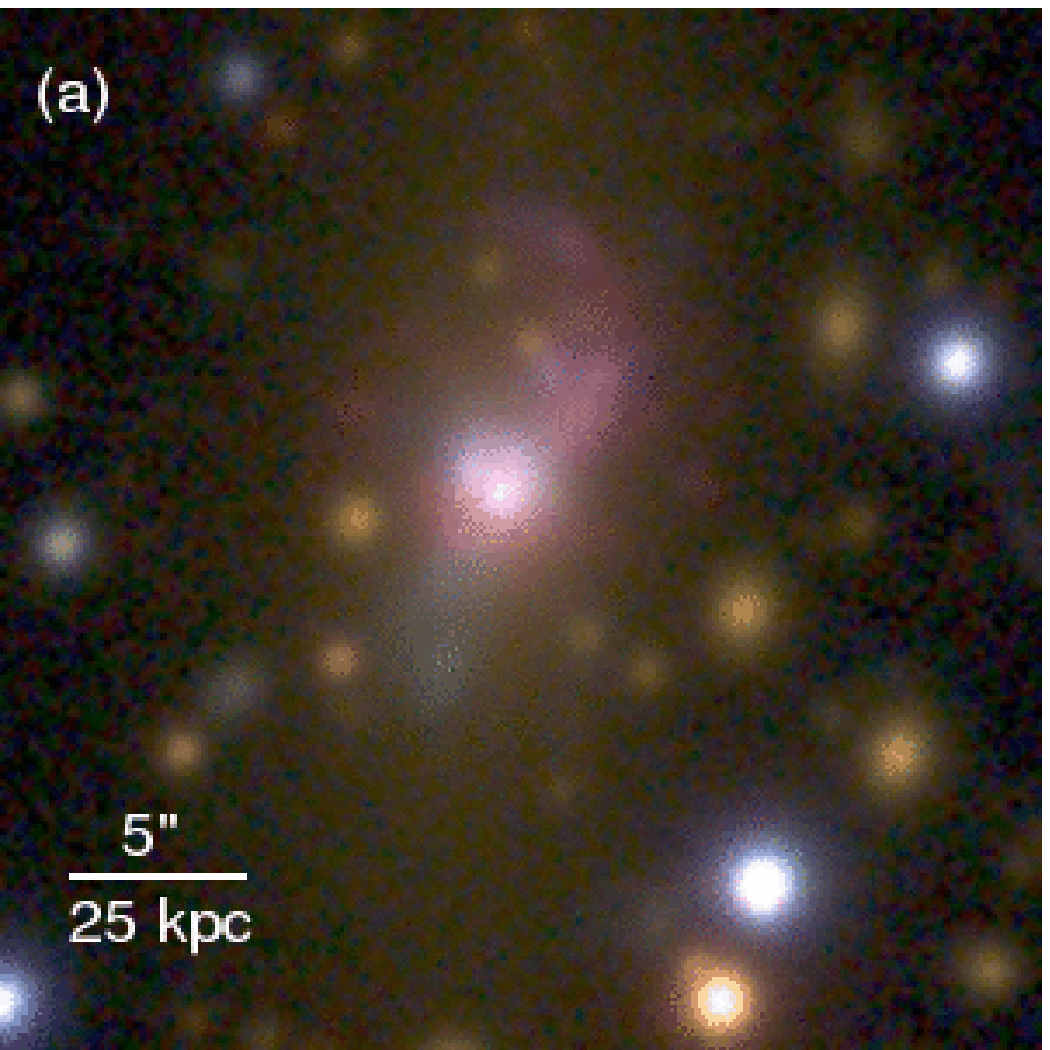}
\includegraphics[width=0.33\hsize]{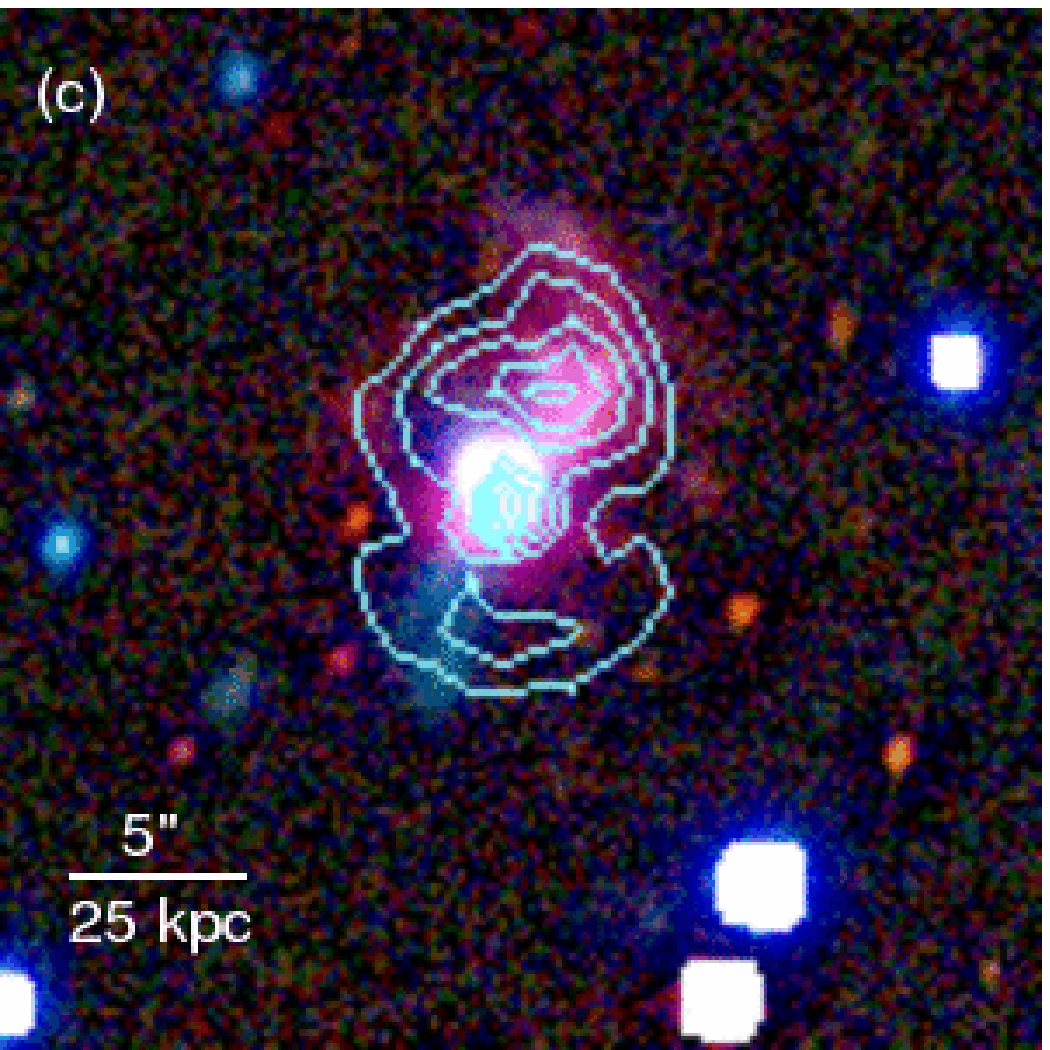}
\includegraphics[width=0.33\hsize]{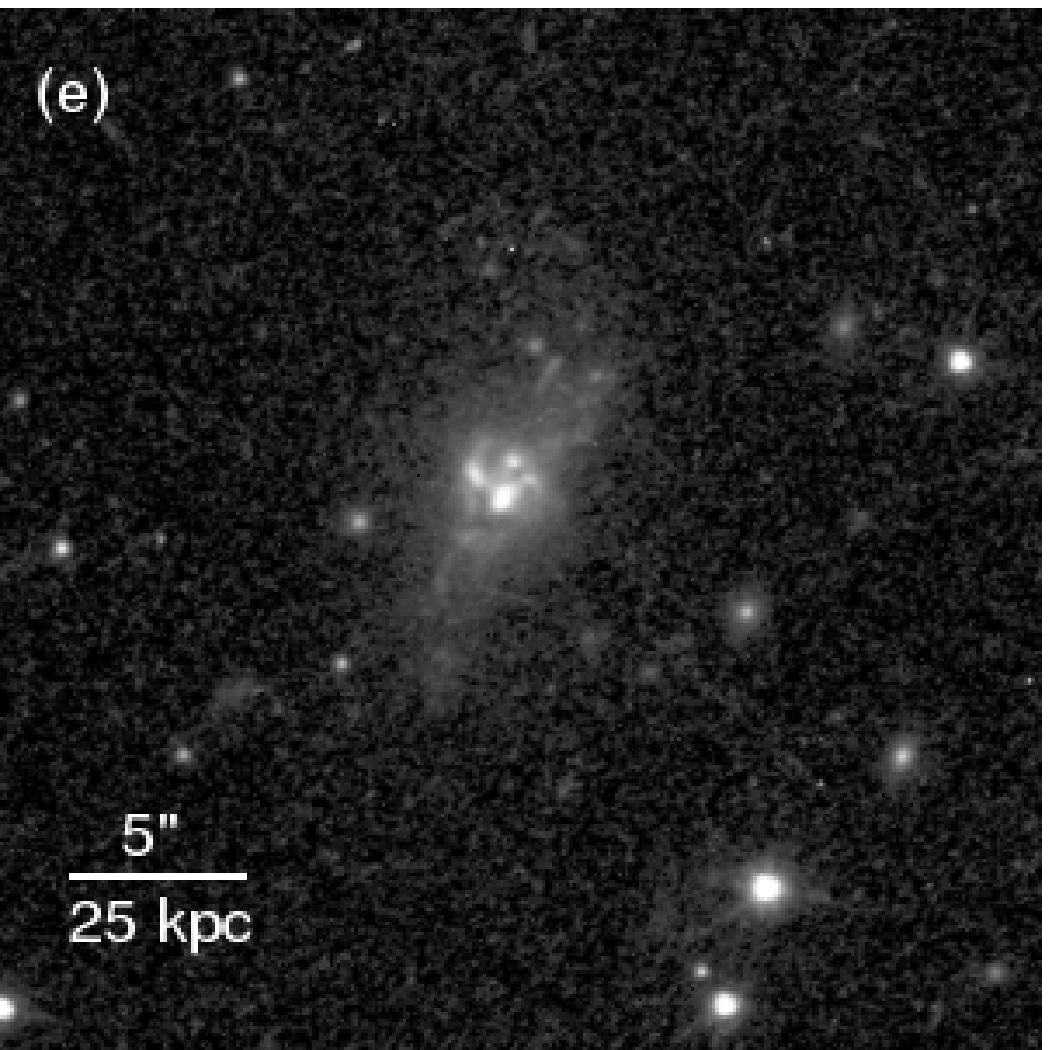}\vspace{0.5mm}
\includegraphics[width=0.33\hsize]{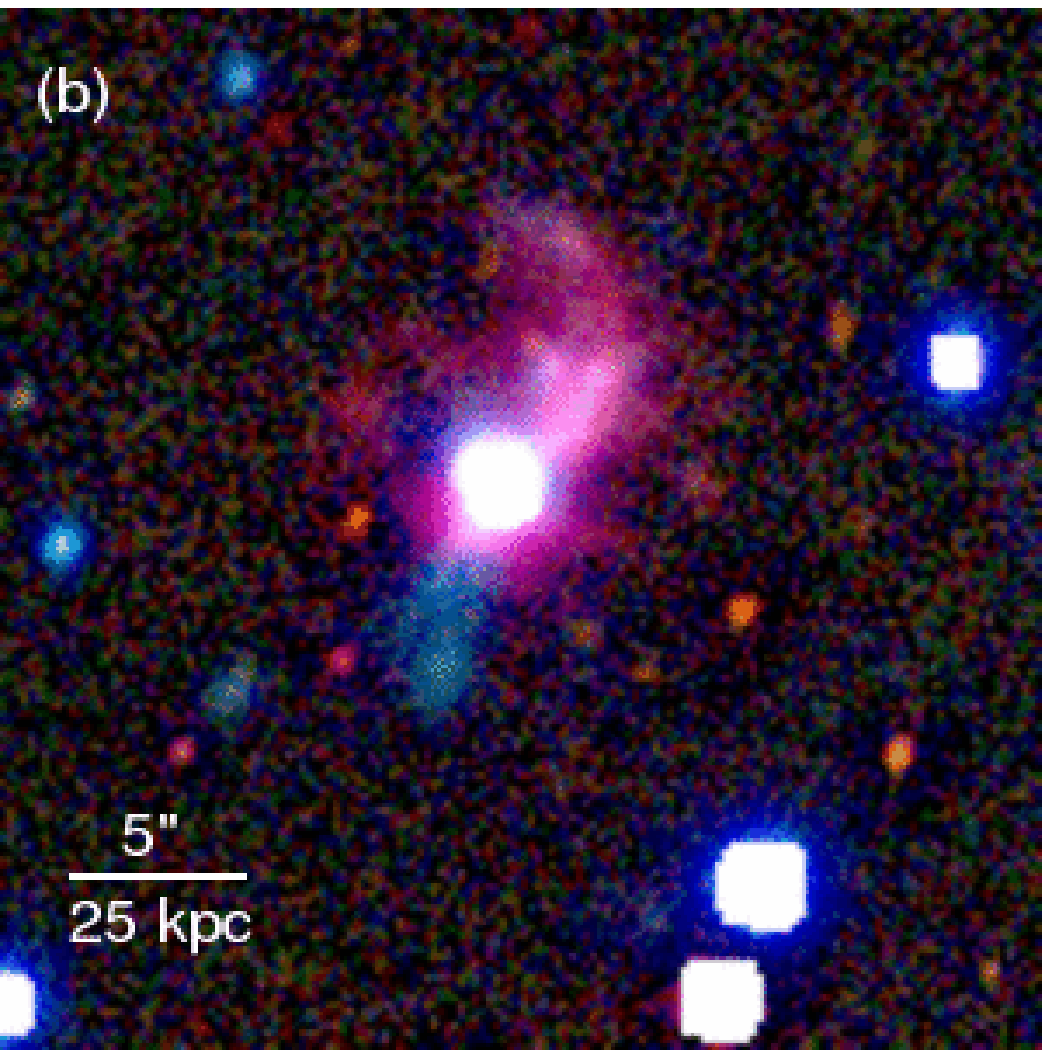}
\includegraphics[width=0.33\hsize]{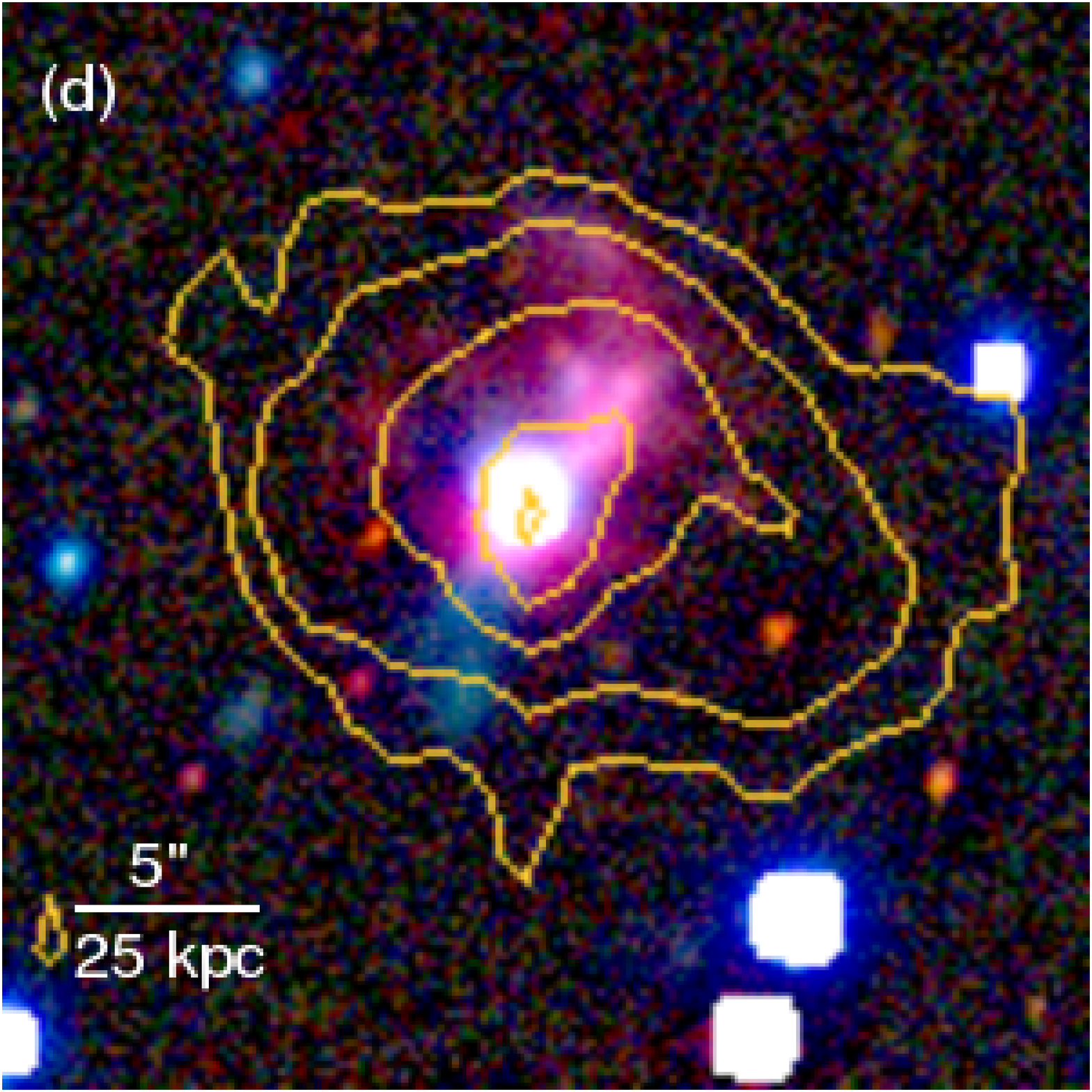}
\includegraphics[width=0.33\hsize]{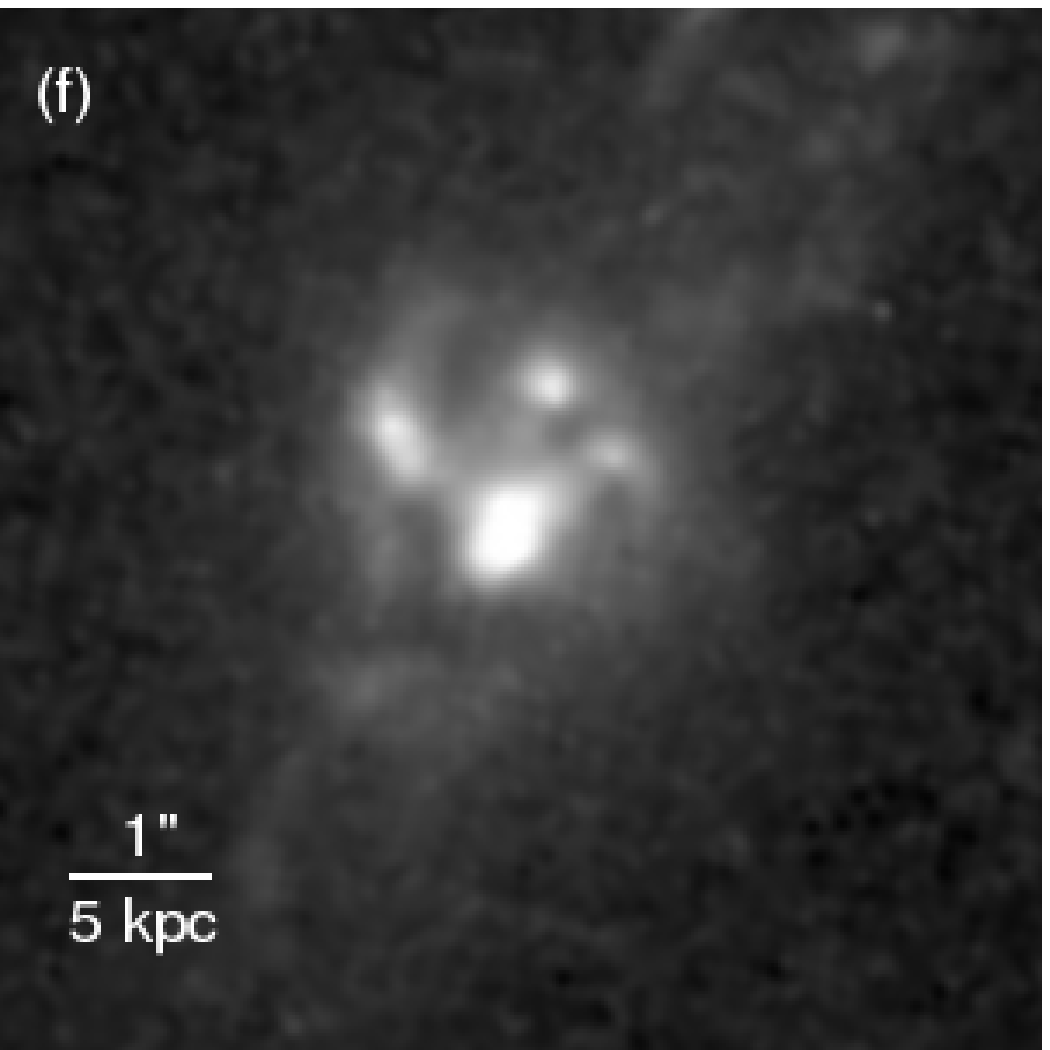}
\caption{Optical structure of the BCG of MACS~J1931.8-2634. (a):
  SuprimeCam {\it BRz} image of the central $30\arcsec \times
  30\arcsec$. (b): For
  this image, the contribution from the old stellar population of the
  BCG (as traced by the SuprimeCam {\it I}-band image) was subtracted
  from each of the {\it B},{\it R} and {\it z} images before combining
  them to a color image. This enhances the blue and pink features
  visible to the southeast and northwest of the central AGN. ``Pink''
  signals contributions from predominantly the blue ({\it B}) and the
  red ({\it z}) channel. At the redshift of the cluster, the H$\alpha$
  line falls into the {\it z}-band, and thus this emission likely
  stems from H$\alpha$ nebulosity surrounding MACS~J1931.8-2634. The
  blue emission, on the other hand, likely signals a young stellar
  population. Interestingly, the H$\alpha$ emission and young stars
  coincide in the northwestern region, whereas in the southeast
  H$\alpha$ emission is absent, or significantly weaker. (c): Contours
  of the X-ray surface brightness map overlaid on the image in
  (b). The brightest knots in the northwestern filament coincide with
  the peak of the cluster X-ray emission north of the AGN
  point source. Thermodynamic mapping shows that this is also the
  coolest, densest part of the ICM, i.e. the cool core. A second peak
  (which also corresponds to cold, dense gas) is also seen close to
  the southwestern filament. The X-ray cavities, on the other hand,
  are located at $\sim$90$\deg$ angles to the bright filaments. (d):
  Overlay of the radio emission on (b).
(e): The HST WFPC2 snapshot, showing the same field of
  view as {\it a)}. (f): The central $7.5\arcsec \times 7.5\arcsec$ of
  the HST snapshot. Note the bright knots in a spiral-like structure
  emanating to the Northwest of the central, brightest knot. }
\label{fig:bcg_zooms}
\end{figure*}

Of particular interest is also the structure of the
BCG. Fig.~\ref{fig:bcg_zooms}(a) shows the central $30\arcsec \times
30\arcsec$ of the SuprimeCam {\it BRz} image.  We see `pink' and
`blue' filaments extending to the northwest and southeast,
respectively.  The `pink' nebulosity to the northwest of the BCG
signals emission in the blue ({\it B}) and red ({\it z}) bands.  At
the redshift of the cluster, the emission lines H$\alpha$, {\sc
  [Nii]}, and {\sc [Sii]} are redshifted into the {\it z} band. The
response of the CCD is not uniform across the filter due to the
decreasing quantum efficiency at longer wavelengths, and H$\alpha$
falls onto the most sensitive wavelength interval of the filter
throughput. In order to single out the line emission, we subtract the
adjacent band ({\it I}) from the {\it z}-band, scaling the {\it
  I}-band image such as to remove the large-scale light distribution
from the dominant stellar population of the BCG. We identify the
resulting emission as predominantly H$\alpha$ line emission, although
this needs to be confirmed spectroscopically.  We also subtract the
{\it I}-band from the {\it B} and {\it R}-bands, in order to better
visualize emission not stemming from the underlying, old stellar
population visible in the {\it I}-band. The resulting image is shown
in Fig.~\ref{fig:bcg_zooms}(b).  The dominant feature here is a bright
filament to the northwest, of both H$\alpha$ emission and blue
emission. At the cluster redshift, the {\sc [Oii]} doublet is shifted
redward of the {\it B} filter, and thus the most likely interpretation
of blue light is continuum emission from young stars. To the
southeast, a filament of bright blue emission is visible, which lacks
strong H$\alpha$ emission. Faint loops of H$\alpha$ and blue light are
also visible to the east and west. Overall, the structure is
reminiscent of NGC~1275, the central galaxy of the Perseus cluster,
which exhibits a rich system of H$\alpha$ filaments, as well as
filaments of young stars \citep{cgw01,fjs08}. From the data at hand,
however, it appears that in MACS~J1931.8-2634, H$\alpha$ filaments are
accompanied by blue filaments (H$\alpha$ alone would be visible as
deep red, not as pink, in our images). The reverse is not always true,
however: we do not detect H$\alpha$ emission coincident with the blue
filament to the southeast.  For comparison, this is significantly
different from NGC~1275, where most H$\alpha$ emission is not
accompanied by star formation \citep{cfj10}.

\begin{figure*}
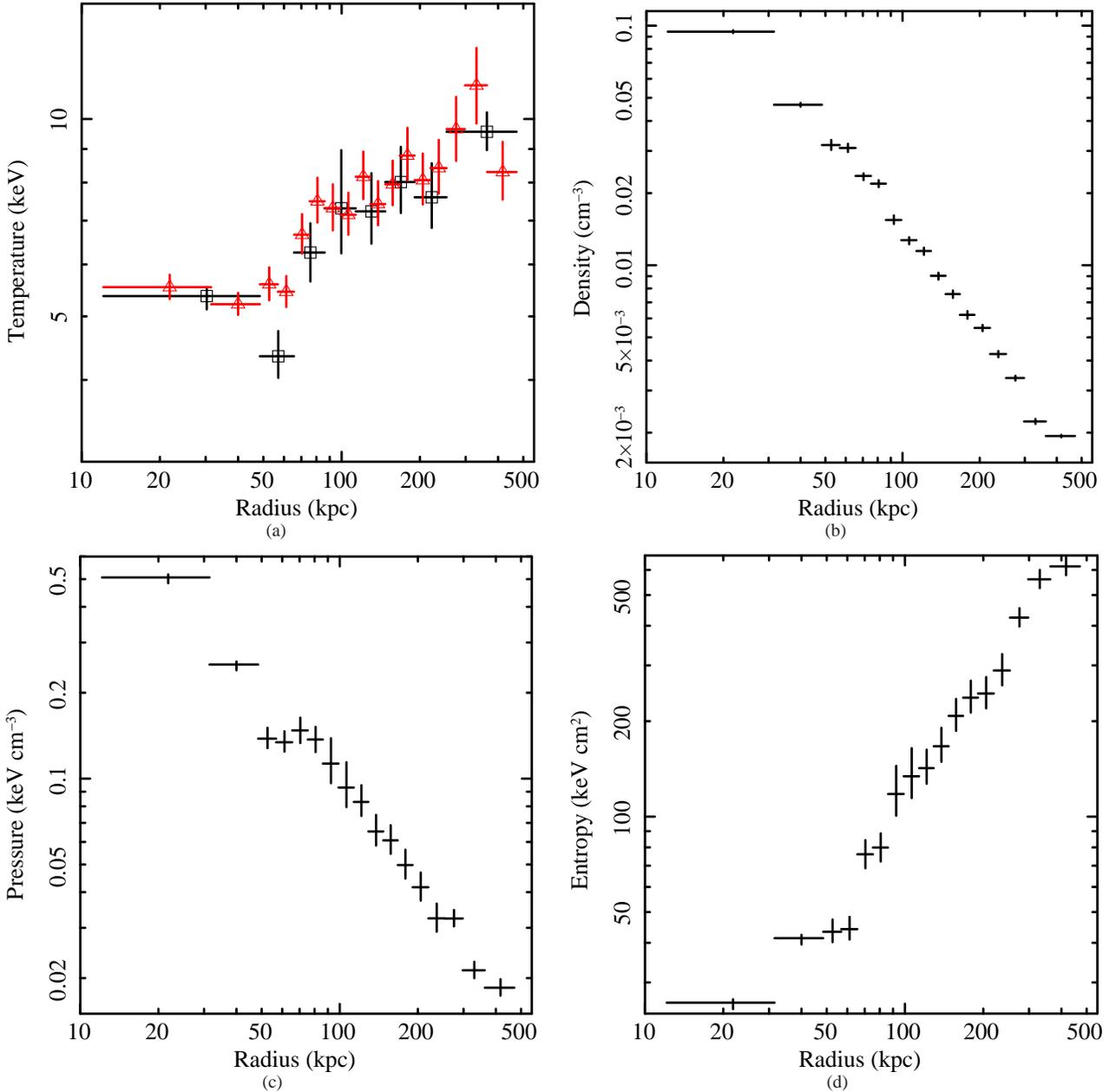

\subfigure[]{
\includegraphics[width=0.45\textwidth, angle=270]{TempNoRidges.ps}
\label{TempProf}
}
\subfigure[]{
\includegraphics[width=0.45\textwidth, angle=270]{DensityNoRidges.ps}
\label{DensityProf}
}
\subfigure[]{
\includegraphics[width=0.45\textwidth, angle=270]{PressureNoRidges.ps}
\label{PressureProf}
}
\subfigure[]{
\includegraphics[width=0.45\textwidth, angle=270]{EntropyNoRidges.ps}
\label{EntropyProf}
}
\caption{Azimuthally averaged thermodynamic profiles for \MACS \ with the bright northern and southern ridges excluded.
a) Projected (red triangles) and deprojected (black squares) temperature profiles. b) Deprojected density profile. 
c) Deprojected pressure profile. d) Deprojected entropy profile. There are clear discontinuities in both the temperature 
and density profiles at distances of $r \sim 70 \kpc$  and further discontinuities in the density profile,
the origin of which may be either bulk motion of cold fronts or weak shock heating.}
\label{ThermoProfs}
\end{figure*}


In Fig.~\ref{fig:bcg_zooms}(c) we overplot contours from the X-ray
emission from Fig.~\ref{NicerImages}(b). We see that the brightest
knots of the northwestern filament (which are bright both in H$\alpha$
and blue light) coincide with the northern ridge identified in the
X-ray imaging and thermodynamic mapping (Sect.~\ref{thermomapsec}). The southwestern
filament also could be associated with the southern ridge. The
X-ray cavities, on the other hand, are at almost $90 \degree$ angles to the
bright filaments.

In Fig.~\ref{fig:bcg_zooms}(d) we overplot contours from the radio
emission (Sect.~\ref{radiodata}). The brighter radio emission seems to coincide
with the brightest H$\alpha$ emission, whereas the fainter radio
emission is elongated in an East-West direction, following the X-ray
cavities.

Finally, in Fig.~\ref{fig:bcg_zooms}(e) and (f) we show the HST WFPC2
snapshot exposure of MACS~J1931.8-2634 (1200s, F606W filter). This
allows us to see details at the very core of the BCG, which are
unresolved in the deeper SuprimeCam images. Apart from the central AGN
point source, we see three bright knots to the north, along
spiral-shaped filaments. 



Giant H$\alpha$ filaments and on-going star formation, as evident by
the presence of young, blue stars, are a common occurrence in the BCGs
of cool-core clusters \citep{all95,cae99,hcf07,rmn08}. In the most
H$\alpha$ luminous systems, the H$\alpha$ luminosity scales well with
the star formation rate \citep{all95, cae99, obp08}. However, in
general, star formation is not the only source of photo-ionizing
radiation in BCGs, as is evidenced by H$\alpha$ emission present even
in the absence of young stars, the spatial offset of H$\alpha$
filaments and young star clusters observed in some near-by systems
\citep{csf05,cfj10}, and the higher {\sc [Nii]}/H${\alpha}$ line
ratios when young stars are not present \citep{hcf07}.  In the BCG of
MACS~J1931.8-2634, young stars are clearly associated with the
H$\alpha$ emission. The H$\alpha$ emission is furthermore coincident
with the ``northern ridge'' defined in Fig.~\ref{RidgeRegions}.  Assuming ratios
of {\sc [Nii]}/H$\alpha$=0.7 and {\sc [Sii]}/H$\alpha$=0.3
\citep[typical of H$\alpha$-luminous BCGs,][]{cae99}, the H$\alpha$
luminosity in this region is $L(H\alpha) \sim 9 \pm 2 \times 10^{42}
h_{70}^{-2}$ erg s$^{-1}$. This would make it the most
H$\alpha$-luminous BCG known.  \citeauthor{cae99} furthermore find
that the typical color excess in H$\alpha$ luminous BCGs is $E(B-V)
\sim 0.3$, albeit with large scatter. Assuming Milky-Way type
extinction, the standard \citet{ken98} conversion between $L(H\alpha)$
and star formation rate \citep[which fits H$\alpha$ luminous BCGs
well, again with large scatter, ][]{cae99,obp08} thus suggests ${\rm
  SFR} \sim 170 M_{\sun} {\rm yr}^{-1}$. This is in remarkable
agreement with the X-ray mass deposition rate of the northern ridge
($\sim 165\pm 60 M_{\odot}{\rm yr}^{-1}$). However, one needs to keep in
mind that the SFR value is only a rough estimate, using only
broad-band imaging, and assuming average values of quantities with
large observed scatter.

%

\section{Radio Observations with \vla}\label{radiodata}
\begin{figure*}
\centering
\subfigure[]{
\includegraphics[width=0.47\textwidth,height=0.53\textwidth]{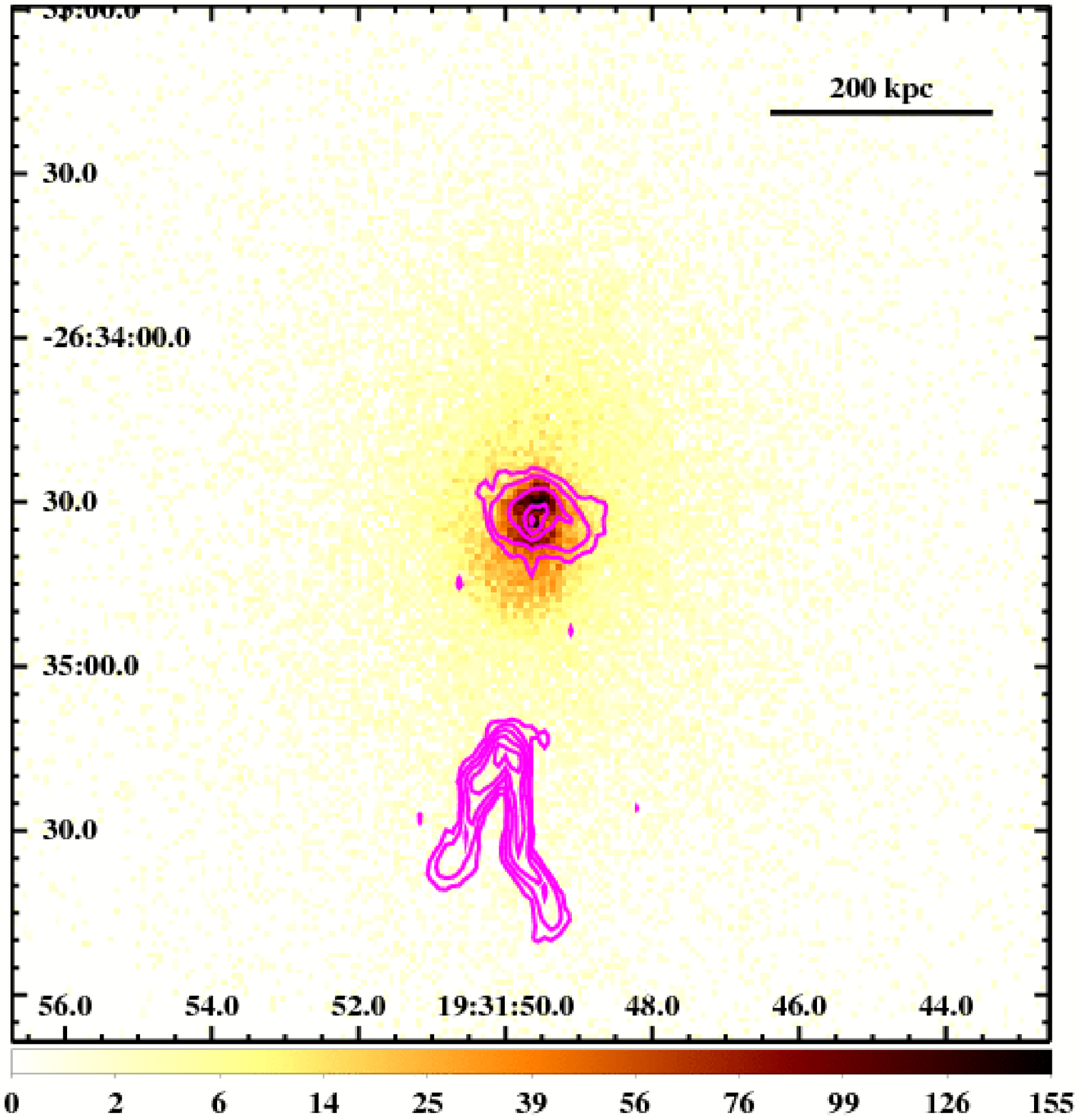}
\label{RadioWide}
}
\subfigure[]{
\includegraphics[width=0.47\textwidth,height=0.53\textwidth]{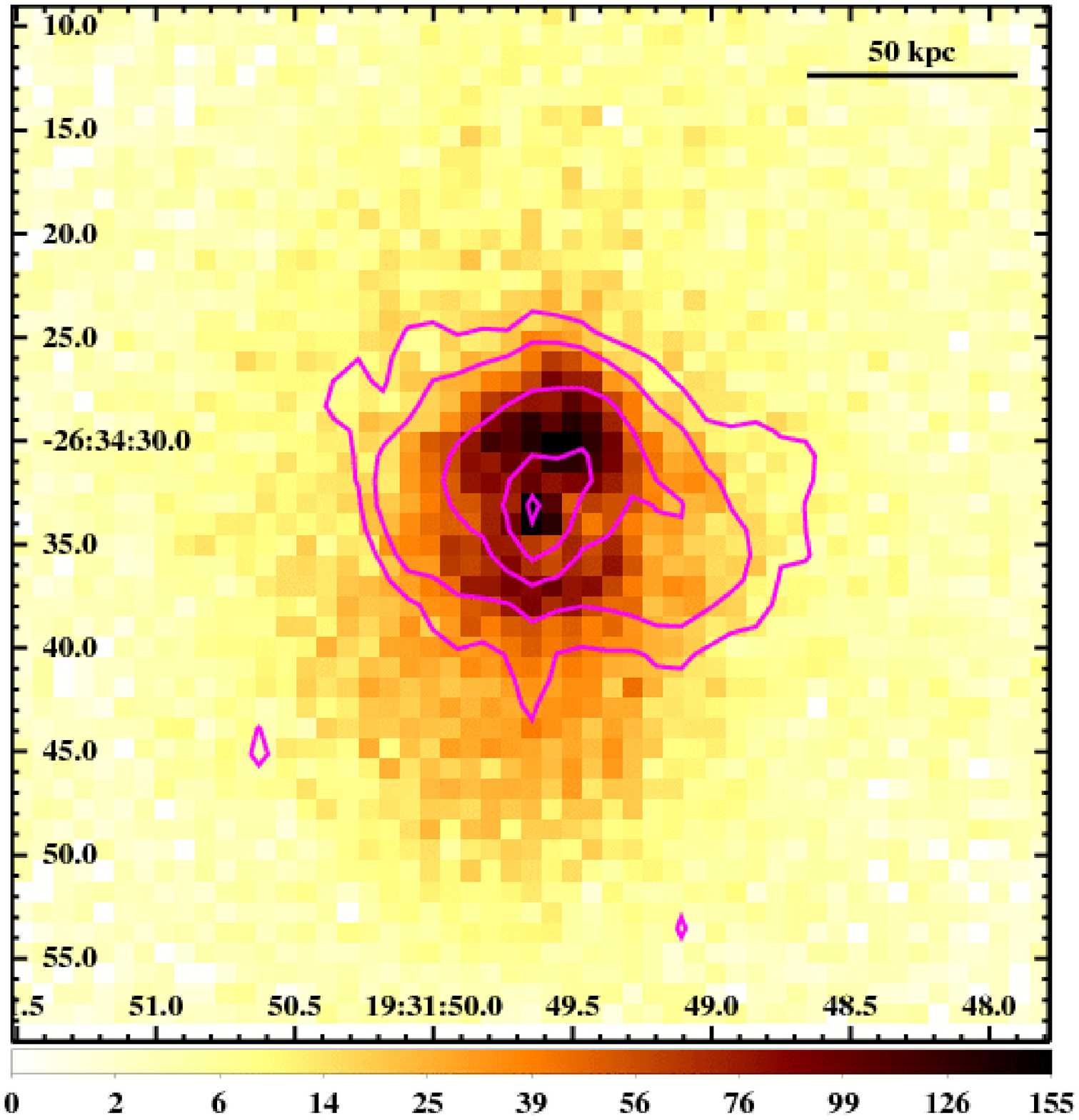}
\label{RadioZoom}
}
\caption{\label{RadioImages} 1.4 GHz radio emission in \MACS \ observed with the \vla. a) X-ray image from Fig. \ref{MACSCont} 
overlaid with radio contours in magenta. The NAT galaxy to south of the center of the cluster is the source of the 
brightest radio emission in this cluster. b) X-ray image from Fig. \ref{MACSZoom} with the radio contours 
overlaid in magenta. This figure shows the amorphous structure of the central radio source, which is clearly centered on the X-ray bright AGN. 
The radio contours are logarithmically spaced between $9 \times 10^{-5} $ and $0.11 Jy$/beam. The beam size for this observation is $1.25 \arcsec \times 2.78 \arcsec$,
and the position angle of the beam ellipse is $5.34$ degrees.
}
\end{figure*}

1.4 GHz radio observations were made with the Very Large Array (VLA) of the National
Radio Astronomy Observatory on 2006 April 14. The data were obtained in A configuration, and the time
on source was 54 minutes.  A central radio source with flux density $\sim$70 mJy is clearly
associated with the central AGN and the core of the elliptical galaxy.  Approximately
45" to the south is a Narrow Angle Tail (NAT) radio galaxy with flux density $\sim$135 mJy.
As shown in Fig. \ref{RadioWide}, the tails of the NAT are swept back to the south, in the same north-south
orientation as the major axis of the central elliptical.  The radio morphology of the central
AGN (seen in Fig. \ref{RadioZoom}) is amorphous without the clearly defined jets or lobes that are found in many radio 
galaxies. Such amorphous radio structures have been seen associated with cD galaxies
in cooling core clusters such as PKS0745-191 \citep{Baum1991,Taylor1994}, 3C317
in A2052 \citep{Zhao1993}, and PKS 1246-410 in the Centaurus cluster \citep{Taylor2002}.
Most likely the radio jets have been disrupted on small scales by dense gas.

\section{The Central AGN: Power and Accretion}\label{centralagn}
The energy of the outburst occurring at the central AGN  manifests itself as both
radiative emission and as a jet that inflates cavities filled with radio
plasma. The energy being input into both of these channels can be estimated from
the observations.

\subsection{Estimating the Radiative Power}\label{AGNRad}
The central AGN is sufficiently bright in X-rays to directly measure
its spectrum and luminosity. Since its spectrum is presumed to be
non-thermal in origin, we have modeled it with a variety of power law models.

\begin{figure*}
\begin{minipage}{0.39\hsize}
\includegraphics[width=0.95\hsize]{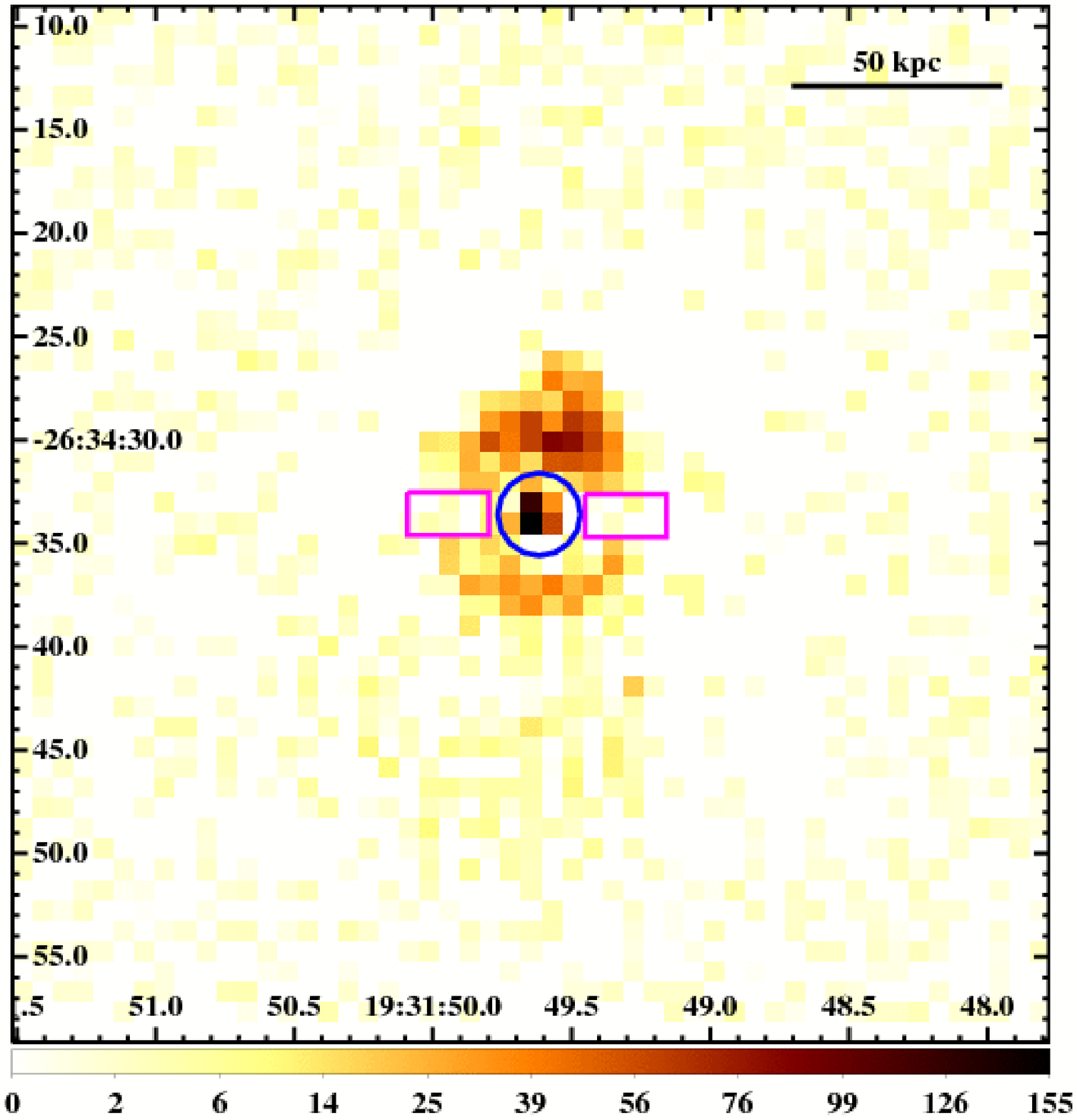}
\end{minipage}%
\hspace{0.015\hsize}%
\begin{minipage}{0.59\hsize}
\includegraphics[width=0.95\hsize]{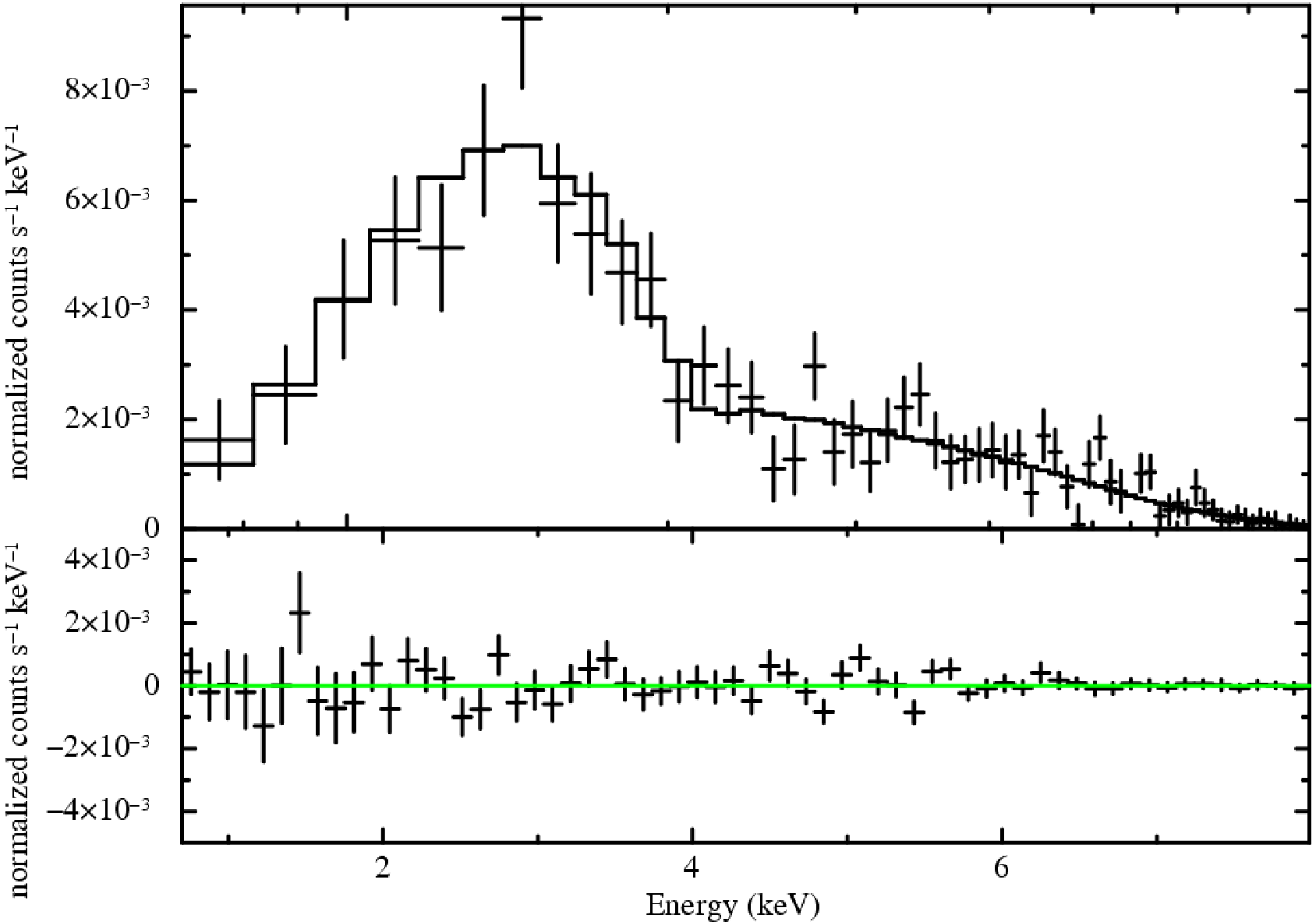}%
\end{minipage}%

\caption{Determining the radiative power emitted from the central AGN. a) The region of spectral extraction for the 
AGN (in blue) and background (magenta). b) The spectrum 
of the AGN with the best-fit power-law model including a fixed Galactic 
absorption and a free intrinsic absorption. The residuals of the fit are shown in the panel below. }
\label{AGNFig}

\end{figure*}

Source counts are extracted from a 2\arcsec \ region centered on the
AGN. The spectral background is extracted from
regions that are chosen to be near the AGN and avoid the 
bright ridges to the north and south, which are likely poor
representations of the true background surrounding the AGN. The source
and background regions are shown in Fig. \ref{AGNFig}(a).
The net spectrum for the AGN using these regions is shown in Fig
\ref{AGNFig}(b). Three different 
assumptions about the absorption have been examined: 1) that the absorption is due only to Galactic contributions, 
and is fixed at the value of \citet{Kalberla2005}; 2) that the absorption
is only due to Galactic contributions, but the column density \mysub{n}{H} is a free parameter in the fit; 
and 3) that the absorption includes both a fixed Galactic component 
and an intrinsic component at the redshift of the cluster with the intrinsic 
column density \mysub{n}{H} a free parameter.
 The results of these fits including the model fluxes and luminosities 
between 0.7 and 8.0 \keV \ are given in 
Table \ref{AGNTable}. It is clear from the fit statistics that a larger absorption column density is favored ($\Delta C >12$, a significant improvement 
at a confidence level well above 99.9\%). 
We conclude that the central AGN has a luminosity in the 
energy range of 0.7-8.0 \keV \ of $\sim$8 $\times 10^{43}$ \ergs. Systematic uncertainties arising from the choice of the background region are estimated at $\sim$30\%.

\subsection{Estimating the Jet Power}\label{AGNJet}
The radio and X-ray emission both suggest that X-ray cavities filled with
radio plasma reside to the east and west of the central AGN. The energy and power required to inflate and fill these cavities ($4PV$) can be 
estimated using the methods described by \citet{Allen2006}, \citep[see also][]{Dunn2004,Dunn2005,Churazov2002}.

\begin{figure}
\centering
\includegraphics[width=0.47\textwidth]{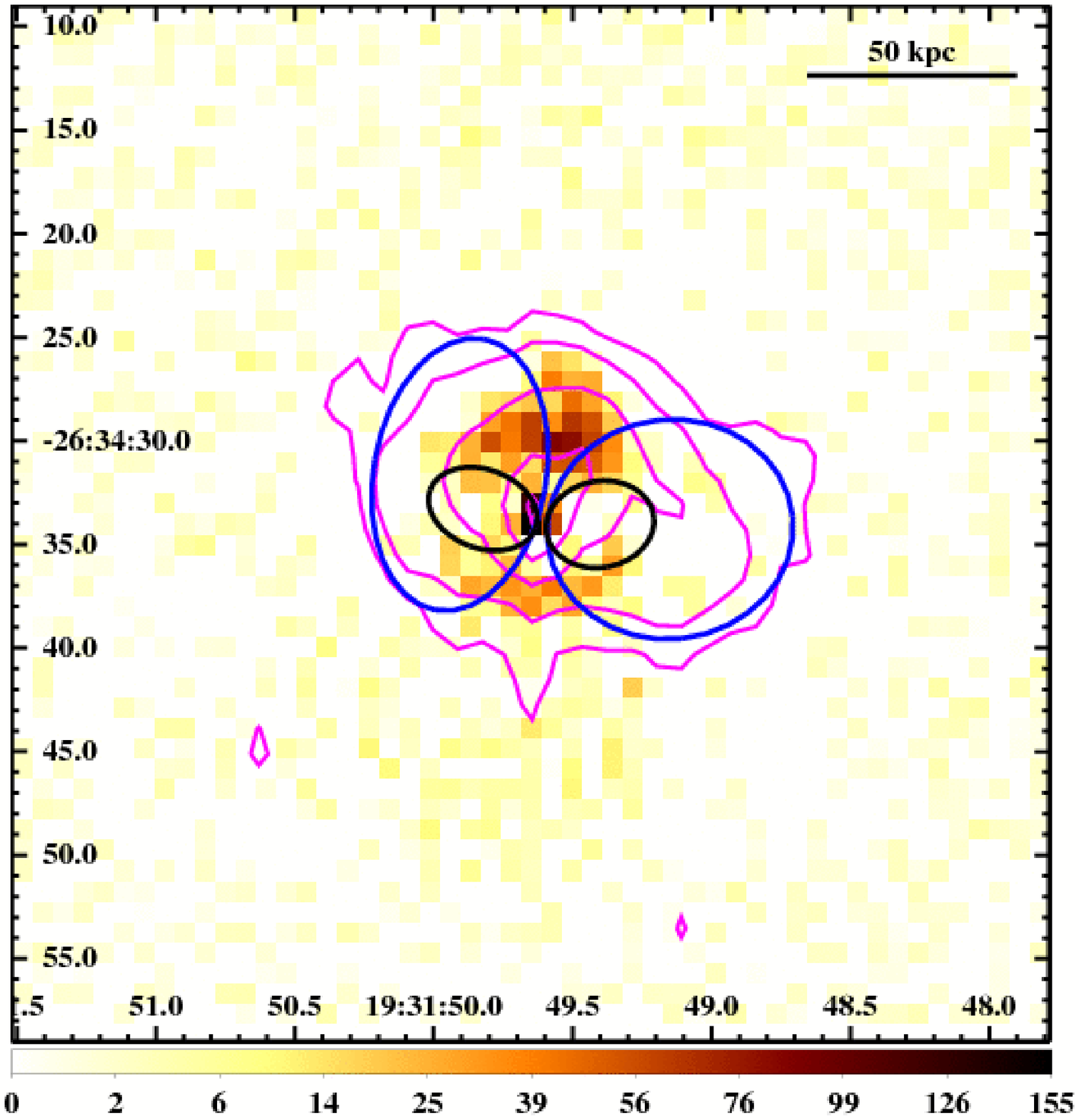}
\caption{\label{Bubbles} Estimating the jet power using the X-ray cavities and radio emission. 
The image of Fig. \ref{Butterworth} is shown with the radio contours overlaid in magenta, the radio cavities overlaid in blue, and the
minimal X-ray cavities overlaid in black.
The determination of the $4PV$ enthalpy 
and jet power are derived from a Monte Carlo analysis that generates plausible elliptical cavities based on these ellipses. The true cavity volumes
are expected to reside between the volumes calculated from these two sets of cavities. }
\end{figure}

Neither the X-ray nor radio image provide an well-defined, unambiguous choice for cavity regions. 
The X-ray image shows two regions to the east and west of the central AGN that 
look like cavities, although the outer boundaries are difficult to determine, particularly given the presence of the bright ridges to the north and south.
In nearby systems, depressions in the X-ray emission associated with the full extent of the radio plasma-filled cavities are typically visible only in 
very deep X-ray observations \citep{Birzan2008}. The radio emission, on the other hand, 
has a much more naturally defined outer boundary, but our 1.4 GHz radio data for this $z=0.35$ cluster do not show clear lobe or jet 
structure to the radio source. The boundaries of this radio emission suggest that it is confined by the surrounding X-ray gas, but we caution that if the radio
plasma were `leaking out beyond' the boundary of the cavities then estimating cavity volumes based on the radio emission would lead to overestimating 
the energy in the cavities \citep{Finoguenov2008}. With these uncertainties in mind, identical calculations were performed with two different sets of cavities:
one set based on the structure of the radio emission (hereafter the radio cavities) and another set based on the apparent cavities in the X-ray images (hereafter the
minimal X-ray cavities). The true cavity volumes, $4PV$ enthalpy, and jet power are expected to lie somewhere between the 
values calculated from these two sets of cavities, both shown in Fig. \ref{Bubbles}. 

Both the radio and minimal X-ray cavities were modeled as tri-axial ellipsoids with volume $V=(4/3)\pi\mysub{r}{l}\mysub{r}{w}\mysub{r}{d}$. 
The measured lengths $\mysub{r}{l}$ and $\mysub{r}{w}$ are the lengths of
the axes in the image plane along and perpendicular to the jet axis, respectively. The final length, \mysub{r}{d}, 
is the axis of the cavity along the line of sight. The initial length of the axis along the line of sight was estimated as the smaller of the two planar axis lengths,
but allowed to vary independently from them.   
Calculating the sound 
speed for the X-ray emitting gas with mean molecular weight $\mu=0.62$ and
adiabatic index $\gamma=5/3$, we estimated the time scale for bubble formation 
as $\mysub{t}{age}=(\mysub{r}{l}/\mysub{c}{s})$. 
From this, we calculated the power required to
inflate the cavities as roughly $\mysub{P}{jet}=(4PV/\mysub{t}{age})$.
A Monte Carlo analysis was performed that drew the temperature,
density, and all three spatial axes from independent Gaussian distributions. The assumed uncertainties on \mysub{r}{l}, \mysub{r}{w}, and \mysub{r}{d} were 
20\%, 30\%, and 30\%, respectively, leading to a systematic uncertainty of $\sim$50\% in the volume. 
Both the temperature and density were calculated from the major axis \mysub{r}{l} using a power-law parametrization of the 
profiles shown in Fig. \ref{ThermoProfs} between 8 and 80 \kpc \ ($X(r)=\mysub{X}{0}r^{\alpha}, X \in kT,\mysub{n}{e}$). 
From these variables the pressure, enthalpy, and jet
power were then calculated. The prior assumptions and subsequent calculations of the cavity energetics are listed in Table \ref{AGNCalcs}.
Although each $4PV$ calculation includes systematic uncertainties 
of the particular cavity volume, the dominant uncertainty is identifying a particular set of cavities, which leads to an uncertainty in the 
$4PV$ enthalpy of approximately a factor of 8 and jet power uncertian within a factor of roughly 4. 
 
We find that the total $4PV$ enthalpy between the two bubbles, after accounting for all uncertainties, is approximately
$1-8 \times 10^{60} \erg$. This corresponds to a power input into the ICM from the jet of approximately 
$\mysub{P}{jet} \sim$ 4 -- 14 $\times 10^{45}$\ergs.
The mechanical energy going into inflating these cavities is nearly two orders of magnitude larger than the
radiative emission of the AGN, and larger than the bolometric luminosity within the central 50 \kpc \ of the cluster. 
Based on the scaling relation of \citet{Cavagnolo2010} and the radio luminosity, the inferred jet power is $\mysub{P}{jet} = 7.7 \times 10^{44}$ \ergs, consistent
with the lower range of jet powers measured here after accounting for the scatter. The temperature profile shows evidence for heating that goes out 
to roughly the same distance as the radio emission, so it is possible that the bubbles are in fact as large as the radio emission even though the X-ray depressions 
are much smaller in scale. The jet power derived from the radio cavities is comparable to the power input measured in the 200 \kpc \ 
cavities of MS0735.6+7421 \citep{McNamara2005}, which was previously the system with the most powerful jets measured.

\section{Discussion}\label{discussion}
As a larger, more luminous, higher redshift analog to nearby systems 
like the Perseus Cluster \citep{Fabian2003,Fabian2006}, \MACS \ provides an extreme 
example of a cluster with a rapidly cooling core and powerful AGN feedback. This powerful AGN outburst combined with 
merger-induced motion has led to a cool core 
undergoing destruction to an extent previously unobserved in galaxy clusters.

There is clear evidence that \MACS \ has undergone a merger event that induced large oscillatory motions of the core. 
On scales of $r \sim$200 $\kpc$, a spiral of cooler, denser gas seen in both the X-ray image and temperature map is observed to wrap
around the core. Such spiral structures arise naturally from mergers and subsequent sloshing and 
are observable for several Gyr after the merger event \citep{Ascasibar2006}.
Clear deviations from elliptical symmetry are seen in the isophotes, whose centroids shift with distance
along the major axis to the north and south. On smaller scales near the core ($r \sim$50 $\kpc$), the X-ray data show that the core has a history 
of motion in the north-south direction. There are two bright ridges to the north and south of the 
central AGN. The northern ridge has a sharp northern edge (possibly a cold front), 
while the southern ridge has a diffuse tail of emission trailing its southern edge.
Both of these features are consistent with the densest gas currently undergoing motion to the north.

\begin{table*}
\caption{\label{AGNTable}Spectral models for non-thermal emission of the  central AGN. 
The three models are described in Section \ref{AGNRad}. 
The absorption column densities are given in units of $10^{22} \cm^{-2}$ and all fluxes 
and luminosities are given in the energy range of 0.7-8.0 \keV and take 
into account their respective absorption.
 The units for the flux are $10^{-13}$ \erg \cm $^{-2}$ \s $^{-1}$, and the units for 
luminosity are $10^{43}$ \erg \s$^{-1}$. Errors listed are $1\sigma$ 
confidence levels. The final column provides the C-statistic and degrees of freedom for the best fit model. }
\centering
\begin{tabular}{ c c c c c c c c } 
\\ \hline {Model} & Galactic \mysub{N}{H} & Intrinsic \mysub{N}{H} & 
{Photon Index} & {Normalization} & Flux & Luminosity & $C/\nu$  \\ \hline\hline 
Fixed Galactic Absorption & 0.083 & 0.0 & $1.24 \pm 0.07$ & $(2.05 \pm 0.15) \times 10^{-5}$ 
& $1.76^{+0.14}_{-0.11}$ & $7.27^{0.58}_{-0.44} $ & 519.81/497\\
\\
Free Galactic Absorption & $0.42 \pm 0.11$ & 0.0 & $1.70 \pm 0.16$ 
& $(3.86^{+0.64}_{-0.96}) \times 10^{-5}$ & $1.97^{+0.18}_{-0.39}$ 
& $8.14^{+0.73}_{-1.63}$& 505.68/496\\
\\
Free Intrinsic Absorption & 0.083 & $0.71^{+0.27}_{-0.22}$ & $1.70 \pm 0.16 $ 
& $(3.83^{+0.88}_{-0.68}) \times 10^{-5}$ & $1.95^{+0.21}_{-0.23}$ 
& $8.05^{+0.89}_{-0.97}$& 506.42/496\\

\hline
\end{tabular}

\end{table*}

\begin{table*}
\caption{\label{AGNCalcs}Calculations of the enthalpy and jet power 
in \MACS \ for the cavities shown in Fig. \ref{Bubbles}.
These calculations follow the procedure described in Section \ref{AGNJet}
and are also discussed in \citet{Allen2006}. The first seven rows are the prior assumptions going into the calculation while the
final six rows are derived quantities.}
\centering

\begin{tabular}{ c c c c c } 
\\ \hline {Parameter} & {East (Radio)} & {West (Radio)} & {East (X-Ray)} & {West (X-Ray)}  \\ \hline\hline 
\mysub{r}{l} (\kpc) & $20.2 \pm 4.0$ & $30.0 \pm 6.0$ & $13.4 \pm 2.7$ & $12.9 \pm 2.6$              \\
\mysub{r}{w} (\kpc) & $25.9 \pm 9.1$ & $24.5 \pm 7.3$ &  $9.4 \pm 2.8$ & $10.3 \pm 3.1$              \\
\mysub{r}{d} (\kpc) & $20.2 \pm 6.1$ & $24.5 \pm 7.3$ &  $9.4 \pm 2.8$ & $10.3 \pm 3.1$             \\
\mysub{kT}{0} (\keV) & $5.30 \pm 0.25$ & $5.30 \pm 0.25$ & $5.30 \pm 0.25$ & $5.30 \pm 0.25$                 \\
\mysub{\alpha}{kT}   & 0 & 0 & 0 & 0            \\
\mysub{n}{e,0} ($\cm^{-3}$) & $3.47 \pm 0.28$ & $3.47 \pm 0.28$ & $3.47 \pm 0.28$ & $3.47 \pm 0.28$             \\
\mysub{\alpha}{ne}     & $-1.17 \pm 0.02$ & $-1.17 \pm 0.02$ & $-1.17 \pm 0.02$ & $-1.17 \pm 0.02$            \\
\\
$V$ ($10^{69} \cm^{-3}$) & $1.25_{-0.50}^{+0.83}$ & $1.64_{-0.65}^{+1.31}$ & $0.11 \pm 0.08$ & $0.14 \pm 0.08$       \\  
\\

$P$ ($\keV \cm^{-3}$) & $0.50^{+0.20}_{-0.10}$  & $0.33^{+0.11}_{-0.06}$ & $0.85 \pm 0.20$ &  $0.81 \pm 0.25$\\
\\
$4PV$ ($10^{60} \erg$) & $4.62 \pm 2.16 $& $3.98^{+2.65}_{-1.51} $ & $0.64 \pm 0.32$ & $0.85 \pm 0.40$ \\
\\
$\mysub{c}{s}$ ($\km \s^{-1}$) & $1170 \pm 30 $ & $1170 \pm 30 $ & $1170 \pm 30 $ & $1170 \pm 30 $ \\
\\
$\mysub{t}{age}$ ($10^{6}$ yr) & $17.0 \pm 3.4 $  & $25.2 \pm 5.02 $ & $11.4 \pm 2.1$ & $10.7 \pm 2.1$\\
\\
$\mysub{P}{jet}$ ($10^{45} \erg \s^{-1}$) & $8.50 \pm 4.61$  & $ 5.36 \pm 3.06$ & $1.90 \pm 1.22$ & $2.42 \pm 1.31$\\ 
\hline
\end{tabular}

\end{table*}

Our optical data independently suggest such a merger: 
the ICL around the cD galaxy is highly elongated in the north-south direction. {\it B}-band emission originating 
from a young stellar population is observed both to the north and
south of the cD galaxy, while \halpha emission originating from ongoing star formation is observed predominantly at positions coincident with the northern ridge. 
The presence of a
young stellar population to the south without any \halpha emission suggests that the primary region of star formation is moving northward.  

In the midst of the motions of the cluster core, a powerful AGN outburst has taken place. 
The central AGN is bright in X-ray emission, with a 
luminosity in the energy band of 0.7-8.0 \keV \ of $\sim$8 $\times 10^{43} $ \ergs. This AGN is surrounded by extended, amorphous 1.4 GHz radio emission. 
The major axis of this radio emission is spatially coincident with depressions in the X-ray emission. The physical extent of the cavities based on the 
observations is unclear, but estimating cavities based on the X-ray and radio emission
gives a robust range for the
$4PV$ enthalpy of the cavities and their corresponding jet power. The $4PV$ enthalpy of these cavities is 
sufficient to counteract the radiative losses from the central 50 \kpc \ for 30-250 Myr. The jet power ($\sim$4 -- 14 $\times 10^{45}$ \ergs) 
identifies \MACS \  among the most powerful cavity sources yet observed. 
The power input into inflating these cavities is approximately 100 times larger than the measured radiative power of the central AGN, and 
four to ten times larger than the bolometric luminosity of the cool core. 

Unlike other more typical cool core clusters \citep{Allen1998,DeGrandi2001,Vikhlinin2005,Pratt2007,Werner2008}, the azimuthally-averaged metallicity profile for \MACS \
shows no significant deviations from a constant value. Assuming that there once 
was a central metallicity peak in \MACS, this suggests that large masses of metal-rich gas has been stripped from the center of the cluster and displaced to the surrounding 
regions. The extent of transport required to account for the flat metallicity profile is strong evidence that the original cool core has 
undergone destructive stripping as it traversed from one side of the cD galaxy to the other.
The only region with an exceptional metallicity is the northern central ridge.
The central metallicity peak in cool core clusters is usually expected to be robust, even in clusters
with powerful central AGN activity \citep{Bohringer2004, Rasera2008}.

Merger and central AGN activity have resulted in the formation of two X-ray bright ridges roughly equal distances north and south of the central AGN.   
The X-ray bright ridge to the north of the central AGN has characteristics usually associated with a cool core. It is the location of the lowest
temperature, highest density gas in the cluster. It is also the location of the most metal rich gas. 
The X-ray spectrum of this northern ridge has a spectroscopic cooling rate of $\dot{M}
\sim$165 \msolaryr down to $0.1 \keV$, in good agreement with the observed star formation rate estimated from the \halpha emission, $\sim$170 \msolaryr. 
This \halpha emission is expected to be almost entirely from star formation, as it is roughly 3 orders of magnitude larger than what expected from Case B Recombination 
\citep{Johnstone1986}. The spectroscopic cooling contributes a large fraction of the emission from this region, estimated at 30\% 
based on the spectral modeling. The northern ridge might therefore be in the early stages of 'catastrophic cooling' \citep[e.g.][]{Fabian1977,Peterson2006}.
The cooler ICM gas to the south of the cD galaxy has a lower metallicity and a current spectroscopic cooling rate consistent with zero, but is also the location of
a young stellar population. This appears consistent with sloshing-induced stripping of the cool core throughout its oscillations along the north-south axis.
The asymmetric thermodynamic structure and different stellar populations of these ridges clearly indicate that there is no longer a single core of low entropy 
gas surrounding the cD galaxy. The extent to which
AGN feedback contributed to the present-day thermodynamic structure of \MACS, apart from core sloshing, is beyond the scope of the current observations. The majority 
of the stripping and disruption of metal rich, low entropy gas from the original cool core could have been caused by the bulk motions,   
but it seems likely that the AGN outburst may have contributed to the separation of the preexisting core into two X-ray bright ridges roughly equal distances
from the central AGN \citep{Guo2010}. This cluster's star formation is exceptional, especially as it does not satisfy all of the empirical conditions for 
central star formation discussed in \citet{Rafferty2008}. Although the cooling time and entropy are below the thresholds put forward by \citet{Rafferty2008},
all of the systems with clear evidence for high star formation rates also have AGN jet powers smaller than the cooling luminosity, which is not the case in \MACS.

Although the extent of stripping and cool core disruption in \MACS \ is substantial, similar phenomena have also been observed in the nearby Ophiuchus Cluster 
\citep{Million2010}. Many of the morphological structures seen in \MACS \ are also seen in the Ophiuchus Cluster. Both clusters have clearly shifting isophotes, inner 
cold fronts, and comet-like diffuse emission trailing to one side of the inner cool core. 
The Ophiuchus Cluster also shows strong evidence for stripped core gas in the form of a metal rich ridge to the north
of the cool core. These processes also appear to be occurring in \MACS, but with the added complications and energy of central AGN feedback.  

The extent to which a cool core can be disrupted or even destroyed by AGN feedback and merger induced oscillations has important 
implications for cosmological studies with clusters \citep[e.g.][]{Burns2008,Mantz2009a,Mantz2009b}. Further observations with X-ray, optical, and 
radio instruments could provide many new insights into the extent that this cool core has been disrupted. Radio observations at higher resolutions 
and lower frequencies could allow for a better understanding of the amorphous central radio source, in particular discerning the origin of the emission by measuring 
the radio spectrum. Higher resolution, deeper observations may also be able to resolve the jet and lobe structure of the central AGN source, which is critical
for further constraining the magnitude and origin of the AGN outburst. Optical
spectroscopy would enable more precise measurements of the star formation rate in the different regions surrounding the
central AGN, and perhaps allow for measurements of the black hole mass. Optical spectroscopy would also allow for measurements of the emission line velocities
within the cool core remnant, providing more details as to the extent of the disruption of the cool core. 
Deeper observations with \cha \ would provide better measurements of the unusual metallicity 
structure and the extent that gas has been stripped from the cool core remnant. The distribution of regions that undergo cooling and the extent of that cooling 
could also be measured in more detail. Finally, a deeper X-ray exposure could shed new light on the thermodynamic structure of the cluster, in particular provide 
more compelling evidence for cold fronts and/or shock heating within the central 100 \kpc. Simulations designed to reconstruct the 
thermodynamic structure of \MACS \ may elucidate the nature of extreme sloshing and feedback, and perhaps also the future evolution of 
such a profoundly disrupted core.

\section*{Acknowledgments}
This work was carried out with \cha \ Observation Awards Number GO8-9119X \& GO0-11139X. 
Support for this work was provided by the National Aeronautics and 
Space Administration through Chandra/Einstein Postdoctoral Fellowship
Award Number PF8-90056 (N.W.) and PF9-00070 (A.S.) issued by the Chandra X-ray 
Observatory Center, which is operated for and on behalf of the National Aeronautics and Space
Administration under contract NAS8-03060. G.G. is a postdoctoral researcher of the FWO-Vlaanderen (Belgium). 
RJHD acknowledges support from the Alexander von Humboldt Foundation.
This research was supported by the DFG cluster of excellence `Origin and
Structure of the Universe' \url{www.universe-cluster.de}.

Further support for this work was provided by the 
Department of Energy Grant Number DE-AC02-76SF00515. The National Radio Astronomy
Observatory is operated by Associated Universities, Inc., under cooperative 
agreement with the National Science Foundation. We also thank 
Silvano Molendi for providing the mean metallicity profile data from \citet{Leccardi2010}. We finally thank the 
anonymous referee for their suggestions and comments that improved this work.

\bibliographystyle{mnras}
\def \aap {A\&A} 
\def \statisci {Statis. Sci.}
\def \physrep {Phys. Rep.}
\def \pre {Phys.\ Rev.\ E}
\def \sjos {Scand. J. Statis.} 
\def \jrssb {J. Roy. Statist. Soc. B} 


%

\def \araa {ARA\&A}
\def \aj {AJ}
 \def \aas {A\&AS}
\def \apj {ApJ}
\def \apjl {ApJL}
\def \apjs {ApJS}
\def \mnras {MNRAS}
\def \nat {Nat}
 \def \pasj {PASJ}
 \def \pasp {PASP}
\def \gca {Geochim.\ Cosmochim.\ Acta}
\def \prd {Phys.\ Rev.\ D}
\def \prl {Phys.\ Rev.\ Lett.}

\bibliography{MACS1931,optical}

\end{document}